\documentclass[aps,prd,floatfix,twocolumn,notitlepage,superscriptaddress,preprintnumbers,nofootinbib,10pt]{revtex4-1}

\usepackage{natbib,setspace}
\usepackage{graphicx}
\usepackage{amssymb}
\usepackage{amsmath,amssymb}
\usepackage{color}
\usepackage{fancyhdr}
\definecolor{darkblue}{rgb}{0,0,0.5}
\usepackage[colorlinks,linkcolor=darkblue,citecolor=darkblue,urlcolor=darkblue]{hyperref}

\allowdisplaybreaks

\newcommand\nn{\nonumber}

\begin{document}
\date{\today}

\title{Collider Constraints on Lepton Flavor Violation in the 2HDM}

\author{R. Primulando}
\email{rprimulando@unpar.ac.id}
\affiliation{Center for Theoretical Physics, Department of Physics, Parahyangan Catholic University, Jl. Ciumbuleuit 94, Bandung 40141, Indonesia}

\author{J. Julio} 
\email{julio@lipi.go.id}
\affiliation{Indonesian Institute of Sciences (LIPI), Kompleks Puspiptek Serpong, Tangerang 15314, Indonesia}

\author{P. Uttayarat}
\email{patipan@g.swu.ac.th}
\affiliation{Department of Physics, Srinakharinwirot University, 114 Sukhumvit 23 Rd., Wattana, Bangkok
10110, Thailand}

\begin{abstract}
In light of the recent CMS analysis on lepton flavor violating (LFV) heavy Higgs searches and updated bounds on various search channels involving neutral and charged scalars, we provide the updated constraints on the Type-III Two-Higgs-Doublet-Model (2HDM) with a $\tau-\mu$ LFV. In doing so, we first extend the CMS analysis to cover the mass region below 200 GeV by recasting their data. After obtaining the bounds on the heavy Higgs production in the mass range between 130 GeV and 450 GeV, we analyze the parameter space of the Type-III 2HDM with various scenarios of mass spectrum and heavy Higgs production strengths. We found that in most scenarios, searching for the heavy Higgs in the mass range lower than $2 m_W$ is very important in constraining the parameter space of the Type-III 2HDM. Hence, we suggest for the future analysis that the search window for the LFV heavy Higgs be extended to the lower mass region.
\end{abstract}

\maketitle

\section{Introduction} \label{sec:intro}
%
The discovery of the 125-GeV resonance ($h$) resembling the Standard Model (SM) Higgs  is arguably one of the most important findings in particle physics. Through a series of measurements, it has been demonstrated that the way the scalar interact with gauge bosons (i.e., $\gamma$, $W$, and $Z$) and third generation fermions is consistent with the SM expectations. 
However, one should note that the present data cannot rule out  the presence of a new mechanism responsible for electroweak symmetry breaking, including the existence of additional scalars that may take place in the process. The effects of new physics can modify the 125-GeV Higgs couplings or so much as induce interactions that are completely absent in the~SM.

The example of the latter is Higgs lepton-flavor-violating (LFV) couplings, responsible for decays like $h\to e\mu$ or $h\to\mu\tau$. This type of processes can also correlate with the low-energy processes, like $\ell_i\to\ell_j\gamma$. In fact, channels with tau lepton in final states, e.g., $h\to e\tau$ and $h\to \mu\tau$, are found to be stronger in constraining the LFV couplings compared to present low-energy $\tau\to e $ and $\tau\to\mu$ conversions, respectively~\cite{Blankenburg:2012ex,*Harnik:2012pb}. Moreover, this particular channel has attracted great interests~\cite{Bjorken:1977vt,*McWilliams:1980kj,*Babu:1999me,*DiazCruz:1999xe,*Han:2000jz,*Kanemura:2005hr,*Giudice:2008uua,*AguilarSaavedra:2009mx,*Davidson:2010xv,*Goudelis:2011un,*Davidson:2012ds,*Celis:2013xja,Kopp:2014rva,*Sierra:2014nqa,*Crivellin:2015mga,*deLima:2015pqa,*Omura:2015nja,*Dorsner:2015mja,*Altunkaynak:2015twa,*Crivellin:2015hha,*Altmannshofer:2015esa,*Botella:2015hoa,*Das:2015zwa,*Arroyo:2013tna,*Das:2015kea,*Mao:2015hwa,*Omura:2015xcg,*Zhang:2015csm,*Sher:2016rhh,*Han:2016bvl,*Banerjee:2016foh,*Huitu:2016pwk,*Lindner:2016bgg,*Gori:2017tvg,*Arroyo-Urena:2018mvl,*Altmannshofer:2018bch,*Hou:2019grj,*Arganda:2019gnv,Buschmann:2016uzg,Primulando:2016eod}
and remains important for constraining many models beyond the SM.  The most recent bound of this $\tau-\mu$ channel, found by using $35.9~{\rm fb}^{-1}$ data at 13 TeV center-of-mass energy, is set by CMS, i.e., BR$(h\to \mu\tau)<0.25\%$ at 95\% CL~\cite{Sirunyan:2017xzt}. A slightly weaker bound is obtained by ATLAS, BR$(h\to \mu\tau)<0.28\%$~\cite{Aad:2019ugc}.

One of the new physics model that can accommodate LFV  is the Type-III Two-Higgs-Doublet-Model (2HDM). In this model, the scalar content of the SM is enlarged by an additional electroweak doublet. As a result, the Yukawa coupling matrices of each fermion type can no longer be simultaneously diagonalized, so flavor-violating interactions naturally arise at tree level.\footnote{In other variants of the 2HDM without tree-level flavor violation, there is a discrete symmetry making only one Higgs doublet couple to one type of fermions~\cite{Glashow:1976nt,*Paschos:1976ay}. These variants, commonly called Type-I or Type-II 2HDM, are extensively studied.} Furthermore, in addition to the $h$, there are more scalars, i.e.,  the heavy CP-even Higgs $H$, the pseudoscalar $A$, and the charged Higgs $H^+$. Each scalar exhibits nontrivial flavor violation at tree level too.  Previous work \cite{Primulando:2016eod} (see also \cite{Buschmann:2016uzg}) has shown that in the context of the current model, the heavy Higgs can have a much larger LFV branching fraction. Hence if the Higgs can be produced in a significant amount, this channel might provide a better opportunity to search  for the LFV interactions. 

Some of the LHC collaborations have recently started to search for LFV decays in this direction. The first LHC search for LFV decays of the new neutral Higgs was performed by the LHCb using 2 fb$^{-1}$ of data at 8 TeV center-of-mass energy~\cite{Aaij:2018mea}. 
The LHCb search places a bound on the production cross-section times the $\tau$-$\mu$ branching ratio ($\sigma\times$BR($\tau\mu$)) of the neutral Higgs between 4 pb to 22 pb over the mass range between 45 GeV and 195 GeV. 
This search is followed by a recent CMS analysis using 35.9 fb$^{-1}$ of data at 13 TeV center-of-mass energy~\cite{Sirunyan:2019shc}. 
The CMS search covers a mass range between 200 GeV and 900 GeV. Their reported bounds are between 51.9 fb to 1.6 fb. 
Even though both analyses cover different mass regions, we expect the CMS search to provide a stronger bound than the LHCb in the case the scalar mass is lighter than 200 GeV. 
This is due to the fact that the CMS analysis employs more data and has higher acceptance.

This work consists of two parts based on the recent CMS results. 
First, we show that the CMS results can be extended to a lower mass region. We recast the CMS results to cover the scalar mass between 130 GeV and 450 GeV. 
Then, we study the implication of recast CMS bounds on the allowed parameter space of the Type-III 2HDM. 
In particular, we will compare the heavy scalar LFV bounds against the $h$ LFV bounds and the bounds for other non-LFV searches.
By comparing the excluded parameter space from all the searches, we will show the importance of the LFV heavy Higgs searches in various regions of the Type-III 2HDM parameter space.

\section{Recasting the CMS Analysis} \label{sec:LHC}

In this section we recast the CMS results to cover the mass region lower than 200 GeV. 
We will see that the lower mass bounds is important for excluding the parameter space of the Type-III 2HDM, especially in the case of heavy Higgs produced with a small cross-section.

\begin{table}[!ht]
\centering
\caption{The cuts employed in the CMS search for heavy neutral Higgs LFV decays~\cite{Sirunyan:2019shc}.}
\label{table:cuts}
\begin{tabular}{||c | c | c | c | c |} 
 \hline
  & $\mu \tau_h, 0 j$ & $\mu \tau_h, 1 j$ & $\mu \tau_e, 0 j$ & $\mu \tau_e, 1 j$ \\ [0.5ex] 
 \hline\hline
 $p_T^\mu$ & \multicolumn{4}{c|}{$> 60$ GeV} \\
 $|\eta^\mu|$ & \multicolumn{4}{c|}{$< 2.4$} \\
 $p_T^e$ & \multicolumn{4}{c|}{$> 10$ GeV} \\
  $|\eta^e|$ & \multicolumn{4}{c|}{$< 2.4$} \\
  $p_T^{\tau_h}$ & \multicolumn{4}{c|}{$> 30$ GeV} \\
  $|\eta^{\tau_h}|$  & \multicolumn{4}{c|}{$< 2.3$} \\
  $p_T^\text{jet}$  & \multicolumn{4}{c|}{$> 30$ GeV} \\
  $|\eta^\text{jet}|$  & \multicolumn{4}{c|}{$< 4.7$} \\
  \hline
 $n^\mu$ & 1 & 1 & 1 & 1 \\
  $n^e$ & 0 & 0 & 1 & 1 \\
  $n^{\tau_h}$ & 1 & 1 & 0 & 0 \\
  $n^\text{jet}$ & 0 & 1 & 0 & 1 \\  
  $n^\text{b-jet}$ & 0 & 0 & 0 & 0 \\  
  $\Delta R (\mu,e)$ & - & - & $<0.3$ & $<0.3$ \\   
  $\Delta R (\mu,{\tau_h})$ & $<0.3$ & $<0.3$ & - & - \\
  $M_T^{\tau_h}$ & $<105$ GeV & $<120$ GeV & - & - \\   
  $\Delta(e,p_T^\text{miss})$ & - & - & $<0.7$ & $<0.7$ \\   
$\Delta(e,\mu)$ & - & - & $>2.2$ & $>2.2$ \\  [1ex] 
 \hline
\end{tabular}
\end{table}

\begin{figure*}[h!]
     \centering
         \includegraphics[width=0.45\textwidth]{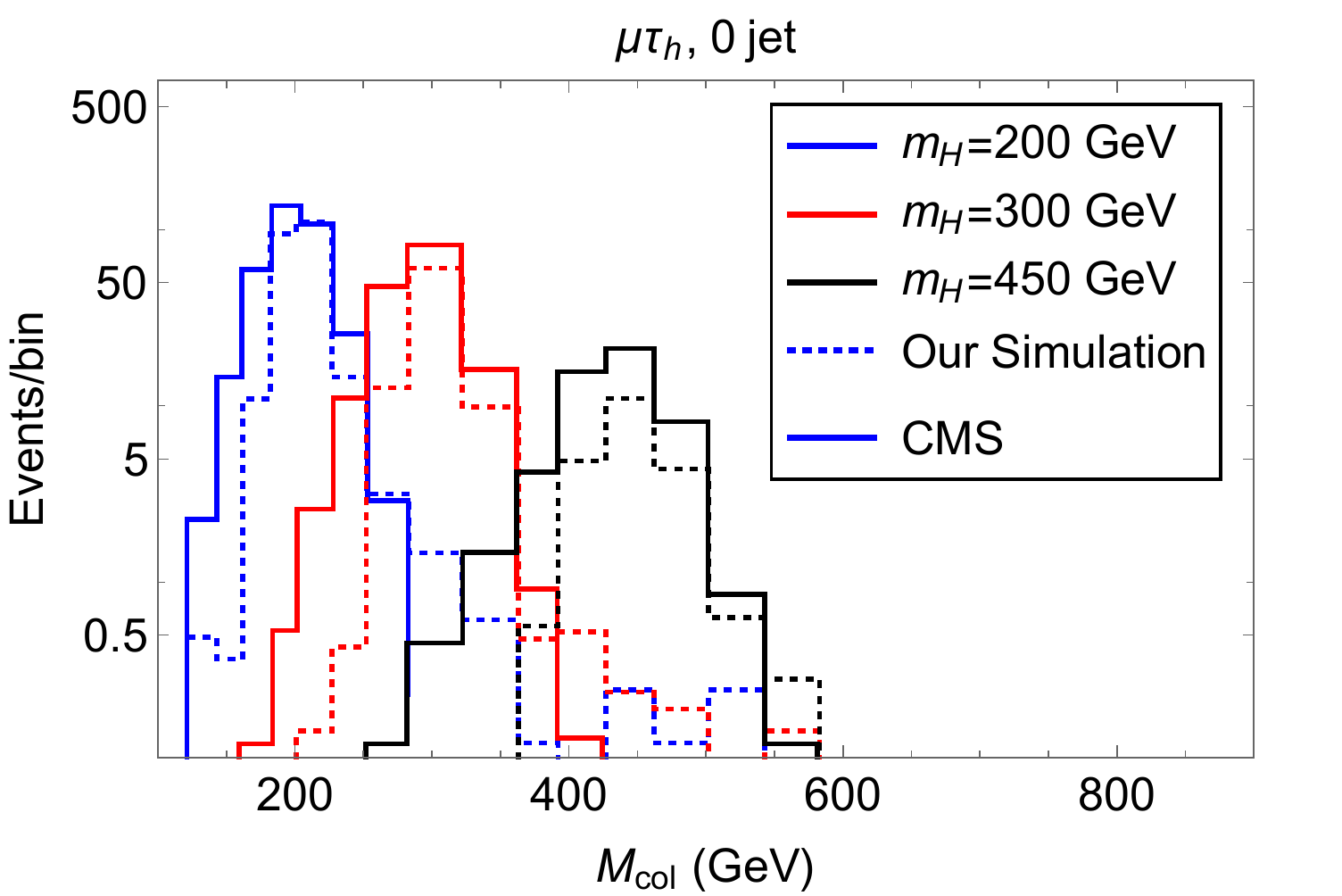}\qquad
         \includegraphics[width=0.45\textwidth]{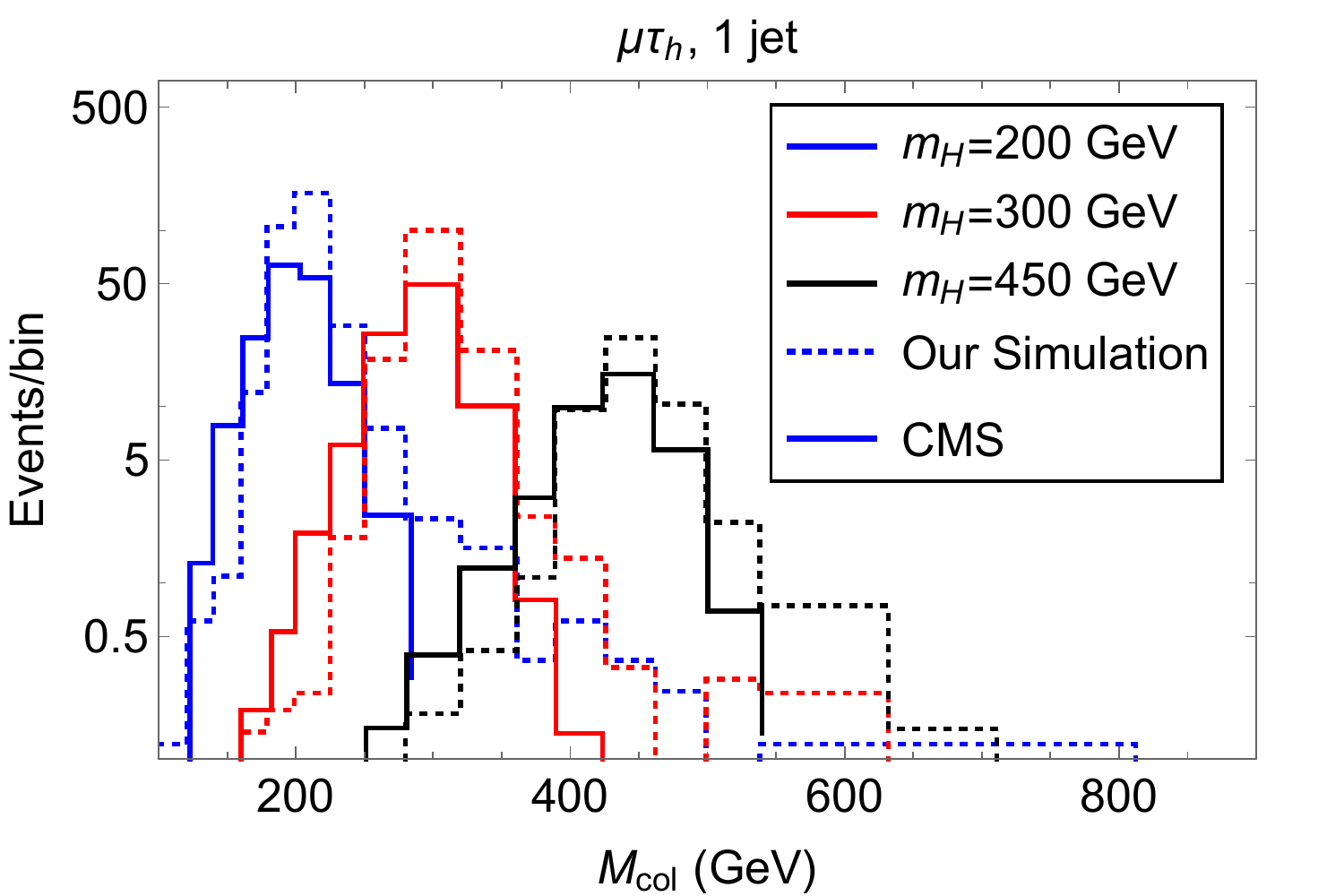}
         \includegraphics[width=0.45\textwidth]{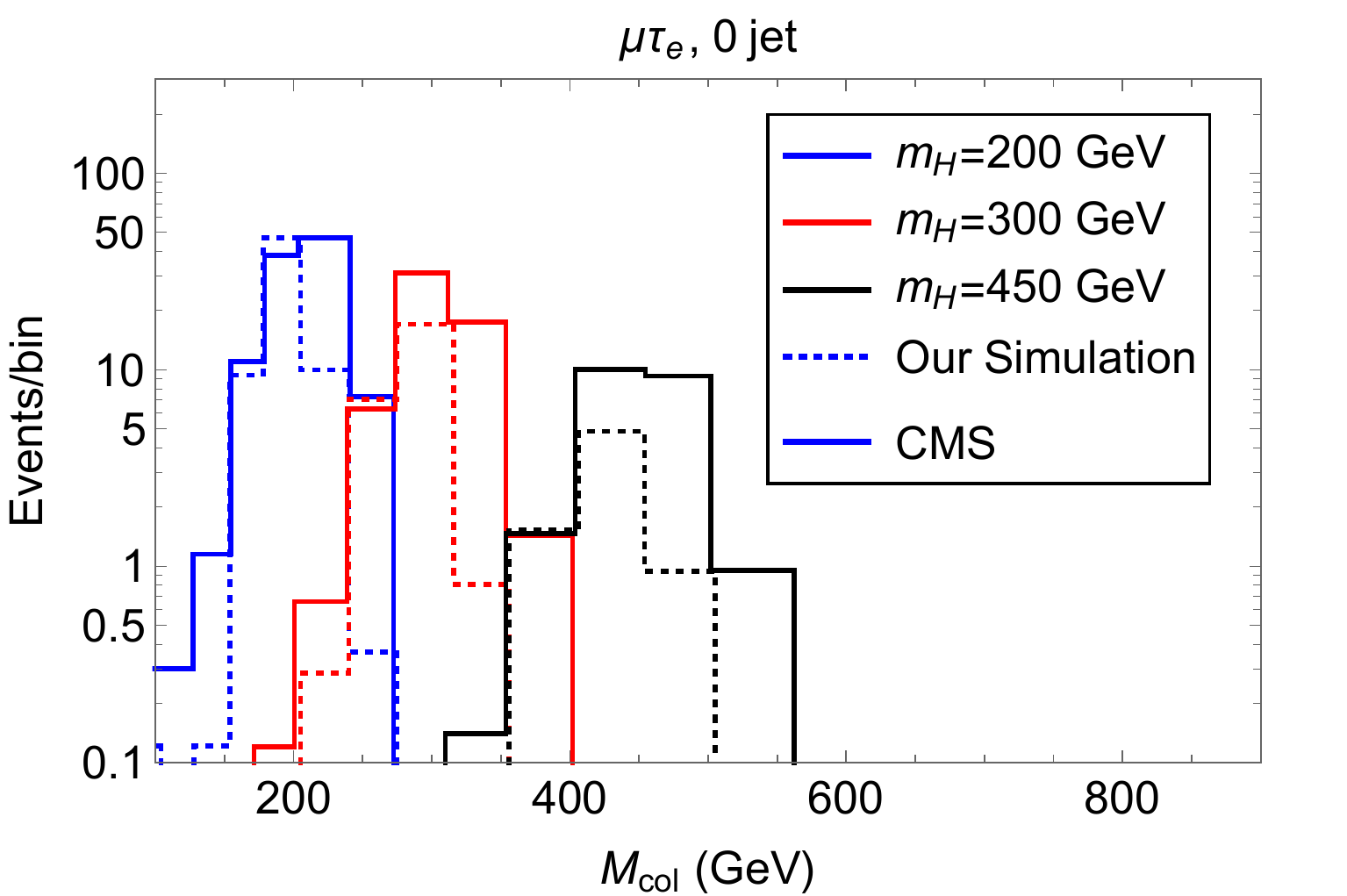}\qquad
         \includegraphics[width=0.45\textwidth]{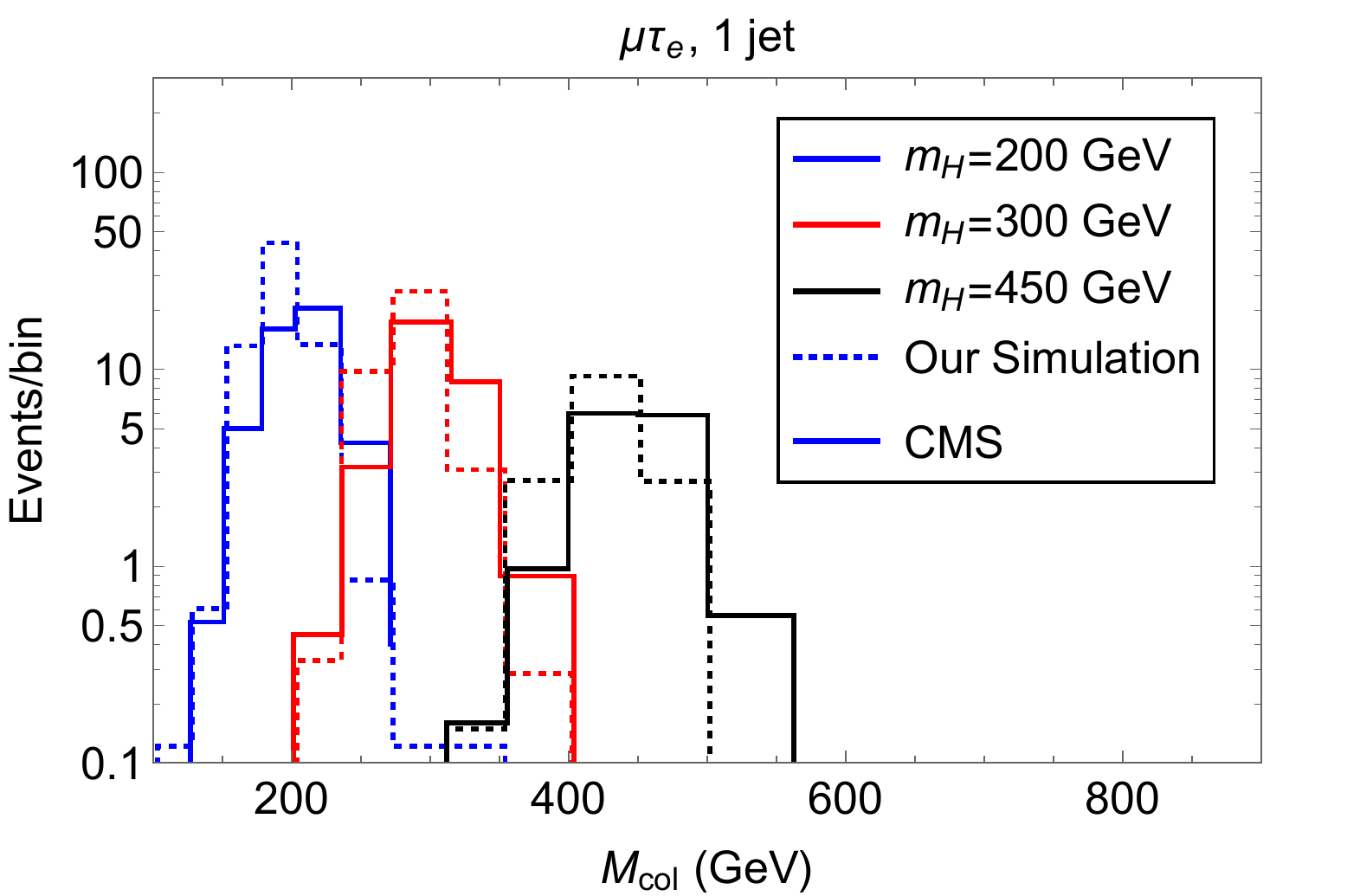}
        \caption{The comparison between our simulation and CMS simulation~\cite{Sirunyan:2019shc} for $m_H =$ 200, 300 and 450 GeV.}
        \label{fig:MC}
\end{figure*}

In order to simulate the signal, we use MadGraph~5~\cite{Alwall:2014hca} followed by shower and hadronization simulation by Pythia~8~\cite{Sjostrand:2014zea}. Detector simulation is done using Delphes~\cite{deFavereau:2013fsa} and we use MadAnalysis 5~\cite{Conte:2012fm,*Conte:2014zja} to analyze the simulated events. The cuts used in the CMS analysis are shown in Table~\ref{table:cuts}. 
The comparison between our monte-carlo simulation and CMS simulation for  $m_H=200$, 300 and 450 GeV is shown in Fig.~\ref{fig:MC}. In the plot, the collinear mass, $M_\text{col}$, is defined as $M_\text{col} = M_\text{vis}/\sqrt{x_\text{vis}}$, where $M_\text{vis}$ is the invariant mass of $\mu-\tau_{h}$ or $\mu-e$. The parameter $x_\text{vis}$ is defined as $x_\text{vis}= \left| \vec{p}_T^{\,\tau_\text{vis}} \right|/\left( \left|\vec{p}_T^{\,\tau_\text{vis}}\right| + \vec{E}_T^\text{miss}\cdot \hat{p}_T^{\tau_\text{vis}}  \right)$, where $\vec{p}_T^{\,\tau_\text{vis}}$ is the transverse momentum of the visible decay of the tau ($\tau_h$ or $e$).

\begin{figure}[h!]
\centering \includegraphics[width=0.45\textwidth]{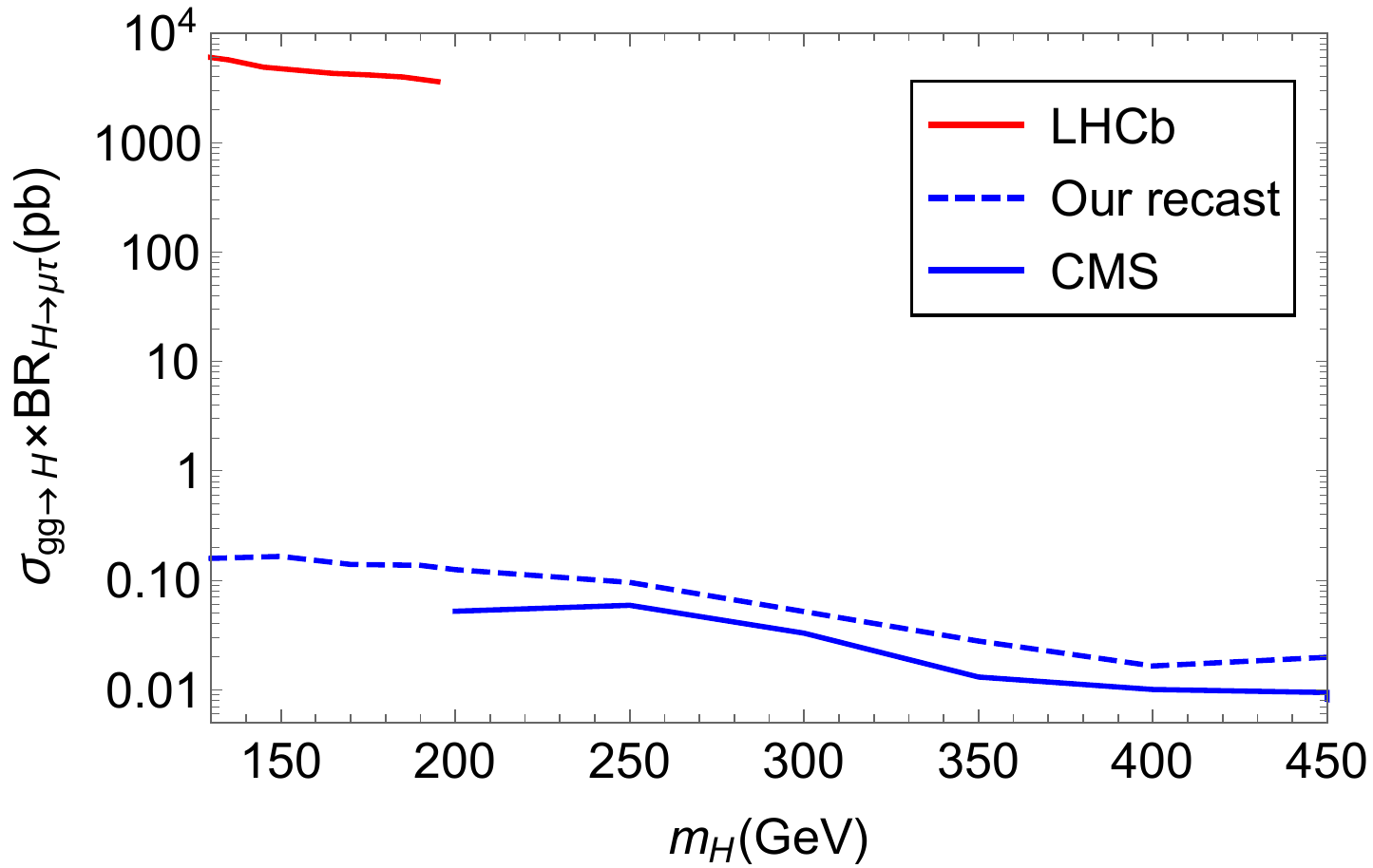}
\caption{The LHC bounds from LHCb~\cite{Aaij:2018mea}, CMS~\cite{Sirunyan:2019shc} and our recast of CMS analysis~\cite{Sirunyan:2019shc}.}
\label{fig:LHCbound}
\end{figure}

In order to derive the bound, we compare the simulated signal and the background. We follow the strategy of~\cite{Balazs:2017moi}  in calculating the likelihood. The likelihood of observing the signal in a particular bin is given by
\begin{equation}
\mathcal L_i \left(n_i | s_i, b_i\right) = \int_0^\infty \frac{\left(\xi \left( s_i + b_i \right) \right)^{n_i} e^{- \xi \left( s_i + b_i \right)}}{n_i!} P_i\left(\xi\right) d\xi,
\end{equation}
where $n_i$ is the number of observed event in the bin, $s_i$ is the predicted number of signal events and $b_i$ is the predicted number of background events. The probability $P_i \left(\xi \right)$ is given by the log-normal function
\begin{equation}
P_i \left(\xi \right) = \frac{1}{\sqrt{2 \pi} \sigma_i} \frac{1}{\xi} \exp\left[ - \frac{1}{2} \left( \frac{\ln \xi}{\sigma_i} \right)^2 \right],
\end{equation}
where $\sigma_i$ is the relative systematic uncertainties of the corresponding bin. The chi squared value is given by
\begin{equation}
\chi^2 = 2 \sum_i \left( \ln \mathcal L \left( n_i | s_i, b_i \right) - \ln \mathcal L \left( n_i | s_i = 0, b_i \right)  \right).
\end{equation}
The 95\% C.L. bounds on the heavy Higgs $\sigma\times$BR($\tau\mu$) for $m_H$ between 130 GeV to 450 GeV are shown in Fig.~\ref{fig:LHCbound}. The cuts utilized by CMS still allow a reasonable acceptance for $m_H < 200$ GeV. Hence from our recast, we show that the CMS data can exclude the region of $m_H$ between 130 GeV to 200 GeV better than the LHCb official analysis, which uses less data. 
Note that our recast analysis gives slightly worse bounds compared to the official CMS analysis for $m_H \geq 200$ GeV, due to the simplistic nature of the statistics we employ. 
In Sec.~\ref{sec:param} we will examine the bounds from the combination of scalar and pseudoscalar productions in our benchmark scenarios.

%
\section{Type-III Two-Higgs-Doublet-Model} \label{sec:2HDM}
%

In this section, we give a brief overview of the Type-III 2HDM. We will closely follow the convention of Ref.~\cite{Primulando:2016eod}, to which the readers may refer to for a detailed description of the model.
\subsection{The Model}\label{subsec:scalarsector}
The hypercharge of both scalar doublets $\Phi_1$ and $\Phi_2$ are taken to be 1/2. 
For convenience in discussing flavor violation, we employ the Higgs basis~\cite{Georgi:1978ri} in which the electroweak vacuum expectation value (vev) resides only in $\Phi_1$. In unitary gauge, the doublet $\Phi_1$ and $\Phi_2$ can be expanded as 
\begin{equation}
	\Phi_1 = \begin{pmatrix}0\\ \frac{1}{\sqrt{2}}\left(v+\phi_1\right)\end{pmatrix},
	\quad
	\Phi_2 = \begin{pmatrix}H^+\\ \frac{1}{\sqrt{2}}\left(\phi_2+iA\right)\end{pmatrix},
	\label{eq:basis}
\end{equation}
where $v=246$ GeV is the electroweak vev. 
The CP-even $\phi_1$ and $\phi_2$ can be rotated to the mass basis by an orthogonal rotation
\begin{equation}
	\begin{pmatrix}\phi_1\\ \phi_2\end{pmatrix}= \begin{pmatrix} \phantom{-}c_\alpha & s_\alpha \\ -s_\alpha & c_\alpha\end{pmatrix}
	\begin{pmatrix}h\\ H \end{pmatrix},
	\label{eq:scalarmixing}
\end{equation}
where $c_\alpha(s_\alpha)$ is a shorthand for $\cos\alpha(\sin\alpha)$. 

The Yukawa couplings of $\Phi_1$ are responsible for fermion mass generation while the Yukawa couplings of $\Phi_2$ induce potential flavor violations. Writing them explicitly, we have~\cite{Primulando:2016eod}
\begin{align}
	\mathcal{L}_{yuk} &= -\bar{L}_L\frac{\sqrt{2}m_\ell}{v}\ell_R\Phi_1 - \sqrt{2}\bar{L}_LY_\ell \ell_R \Phi_2 \nn\\
	&\quad -\bar{Q}_L\frac{\sqrt{2}m_U}{v}u_R\tilde\Phi_1 - \sqrt{2}\bar{Q}_LY_U u_R \tilde\Phi_2 \label{eq:yuk}\\
	&\quad -\bar{Q}_LV\frac{\sqrt{2}m_D}{v}d_R\Phi_1 -  \sqrt{2}\bar{Q}_LVY_Dd_R \Phi_2,\nn
\end{align}
where $m_f$ are the diagonal fermion mass matrices, $Y_f$ are the Yukawa coupling matrices, $\tilde\Phi = i\sigma^2\Phi^\ast$, $V$ is the Cabibbo-Kobayashi-Maskawa matrix and we have suppressed the flavor indices. Note in the above equation all the fermions are in the mass eigenstates with 
\begin{equation}
	L_L = \begin{pmatrix}\nu_L \\ \ell_L\end{pmatrix},\qquad
	Q_L = \begin{pmatrix}u_L \\ V d_L\end{pmatrix}.
	\label{eq:fermiondoublet}
\end{equation}

In the rest of this paper, for brevity, we will refer to the $h$ as the light Higgs and the $H$ as the heavy Higgs.
We will also refer to $H$, $A$ and $H^+$ collectively as the heavy scalars.

\subsection{Productions and Decays of Heavy Scalars}\label{subsec:production}

The light Higgs couplings to SM particles have been measured extensively at the LHC~\cite{Khachatryan:2016vau,*Sirunyan:2018koj,*Aad:2019mbh}. They are consistent with the SM expectations. Thus one would expect the mixing angle $\alpha$ to be small, see Eq.~\eqref{eq:scalarmixing}.  The small mixing angle has a significant impact on the production of the heavy scalars as we will now discuss.

\begin{figure}[h!]
\centering
	\includegraphics[width=0.475\textwidth]{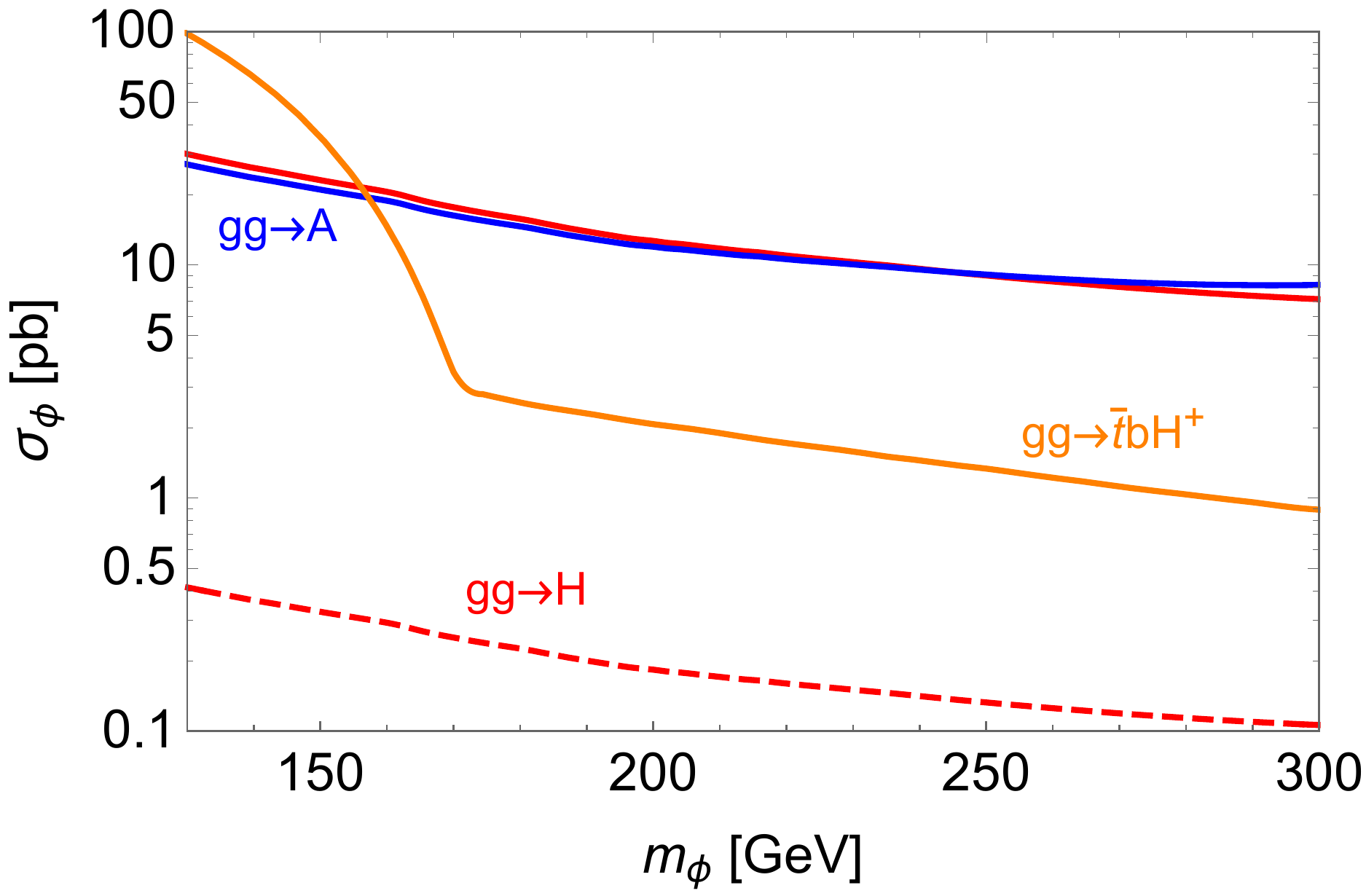} 
\caption{The production cross-sections of the $H$ (red), the $A$ (blue) and the $H^+$ (orange) for $s_\alpha=0.1$ with $Y_U^{tt} = 0$ (dashed lines) and 0.5 (solid lines).}
\label{fig:xsec}
\end{figure}

\begin{figure*}[t!]
\centering
	\includegraphics[width=0.32\textwidth]{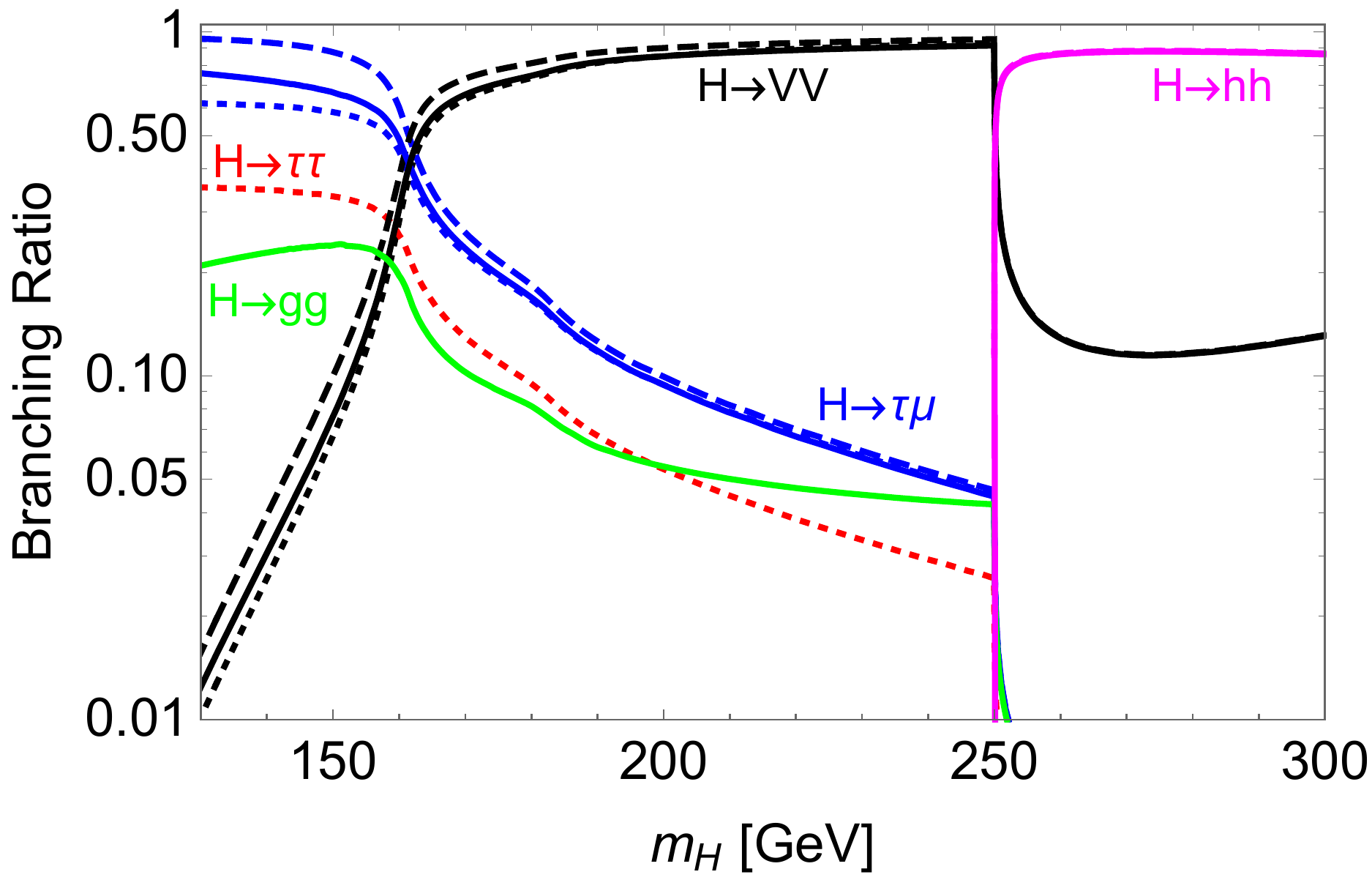} 
	\hspace{.1cm}
	\includegraphics[width=0.32\textwidth]{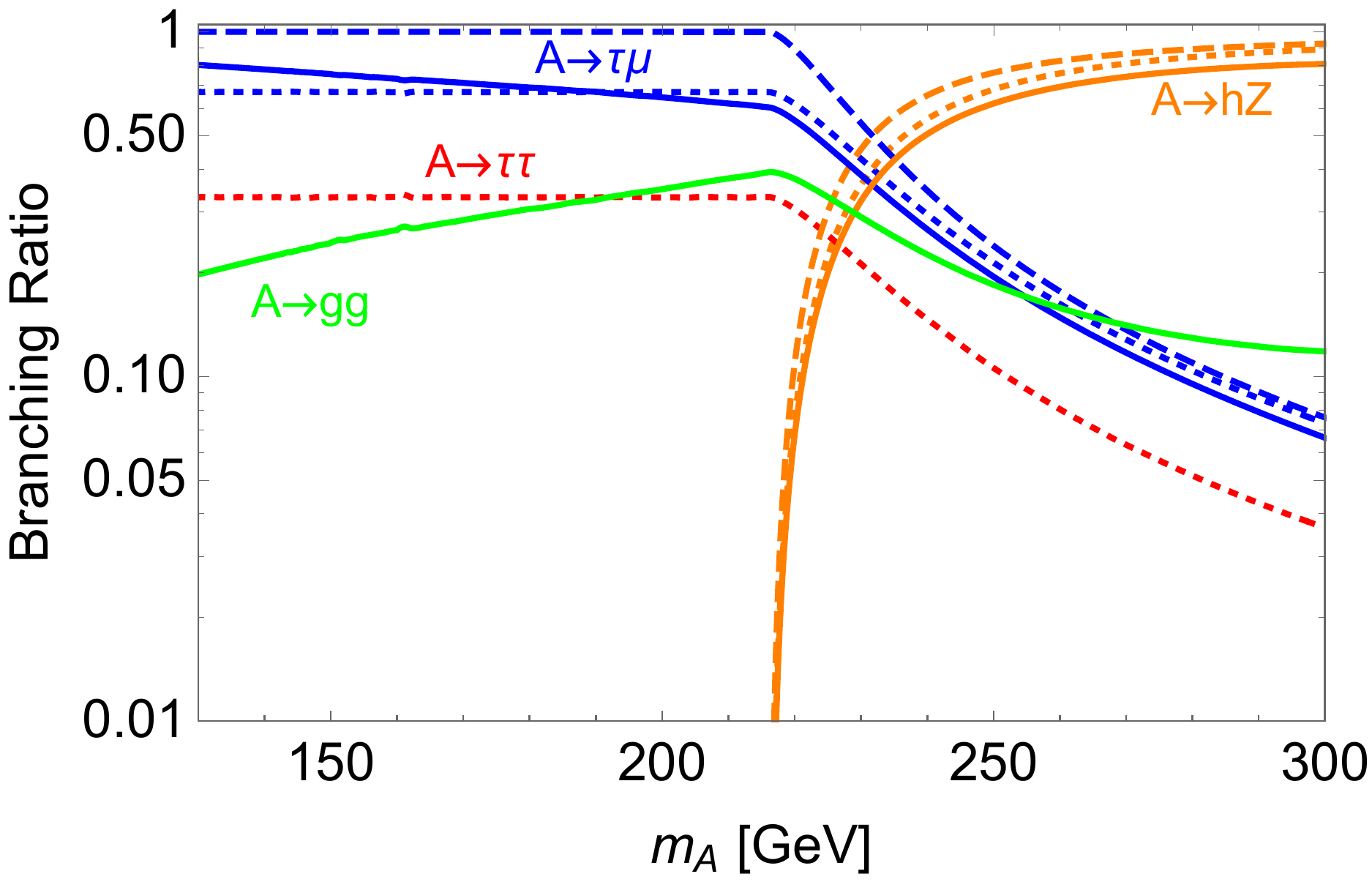} 
	\hspace{.1cm}
	\includegraphics[width=0.32\textwidth]{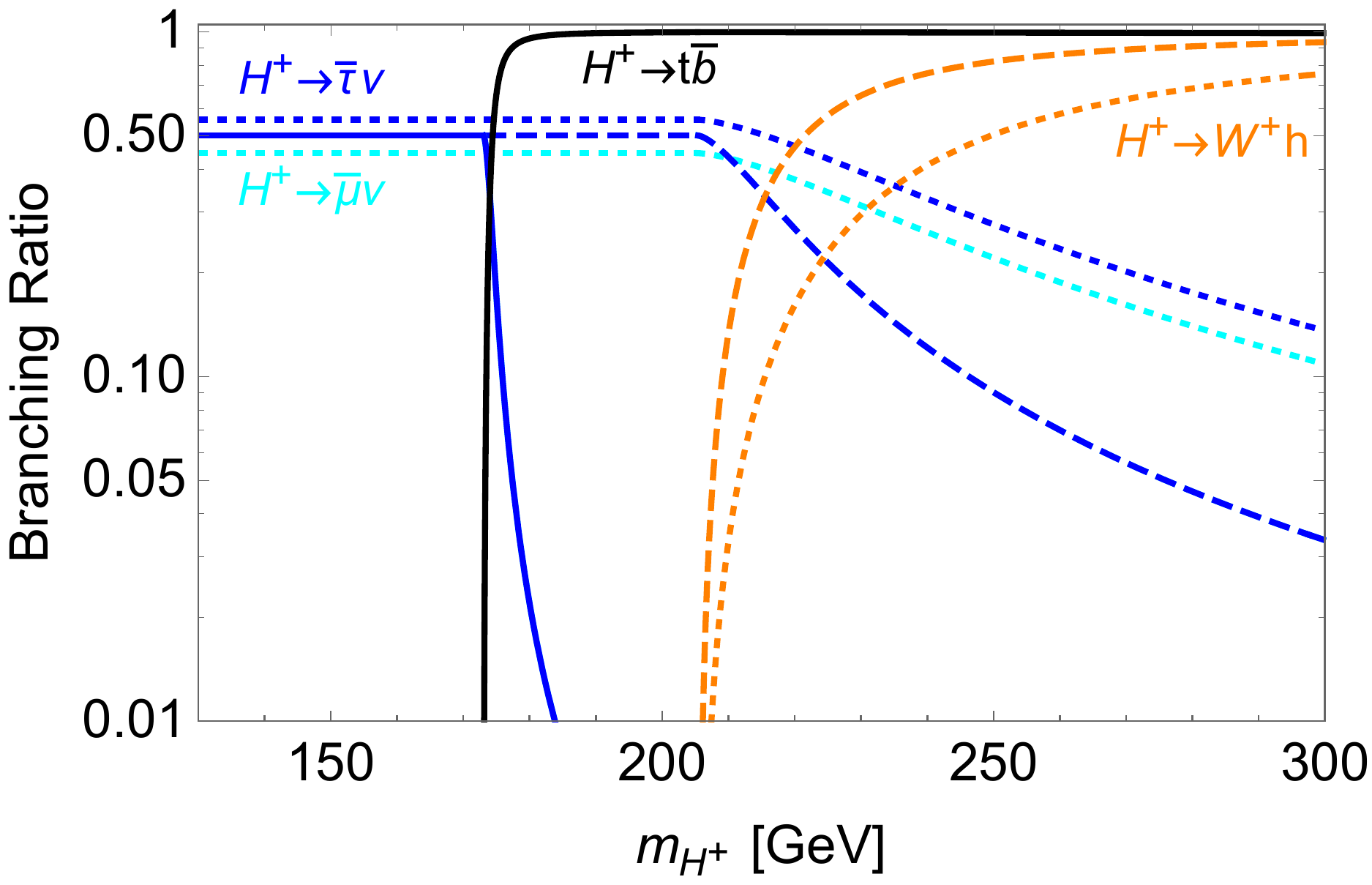} 
\caption{The branching ratios of the $H$ (left) and the $A$ (middle) and the $H^+$ (right) for $s_\alpha=0.1$, $Y_\ell^{\mu\tau}=Y_{\ell}^{\tau\mu}=0.01$ with $Y_U^{tt}=0.1$ (solid lines), $Y_U^{tt}=0$ (dashed lines) and  $Y_U^{\tau\tau} = 0.01$ (dotted lines). Other Yukawa couplings not explicitly mentioned are taken to be 0.}
\label{fig:br}
\end{figure*}

The production cross-section of the $H$ and $A$ are  given in Ref.~\cite{Primulando:2016eod}. The production channels of the $H$ are similar to those of the light Higgs. On the other hand, the $A$ is mainly produced by gluon fusion and associated production with a pair of top quarks while the $H^+$ is produced in association with a top quark and a bottom quark.
Here we note that the $\Phi_2$ Yukawa couplings, especially $Y_U^{tt}$, allow for sizeable production cross-sections of the $H$, the $A$ and the $H^+$.\footnote{The presence of any of the $Y_U$'s or $Y_D$'s would enhance the $H$ production. However, the light quark Yukawas are strongly constrained by dijet searches~\cite{Aaboud:2018zba,*Aaboud:2019zxd,*Sirunyan:2019pnb} and the 125 GeV Higgs data~\cite{Aaboud:2018fhh,*Sirunyan:2019qia}. Moreover, light quark Yukawas would dilute the $H$ LFV branching ratios. Therefore, we choose to consider only the $Y_U^{tt}$ coupling.} 
In the absence of such couplings, a single-scalar productions of $A$ and $H^+$ vanish. However, the single-scalar production of $H$ can still proceed via the mixing in a neutral component of $\Phi_1$. Nevertheless, this production cross-section is suppressed by $s_\alpha^2$. Thus we will loosely refer to the scenario where $Y_U^{tt}$ is present as a large production cross-section case. Similarly, we will loosely refer to the case where $Y_U^{tt}=0$ as the small production cross-section scenario. The production cross-sections of the heavy scalars for both cases are shown in Fig.~\ref{fig:xsec}.

The partial decay widths of the heavy neutral scalars are also given in Ref.~\cite{Primulando:2016eod}. We note here that in the absence of the $\Phi_2$ Yukawa couplings (the $Y$'s), the $H$ would have similar decay channels as the light Higgs, $h$, with additional decay channels to a pair of scalars if they are kinematically open. These extra decay channels arise from the scalar cubic couplings of the $H$ which can be parametrized as $\frac12v\lambda_{H\phi\phi}H|\phi|^2$. Meanwhile, if the $Y$'s couplings are absent, the pseudoscalar and the charged Higgs can only decay through gauge couplings. In particular, possible decay channels for the pseudoscalar in this case include $A\to h(H)Z$, $H^\pm W^\mp$, while possible decay mode of the charged Higgs is $H^\pm \to h(H,A)W^\pm$. 

The LFV decays of the heavy scalars are mediated by the $Y_\ell$ couplings. In this work, we will focus on flavor violation in the $\tau$-$\mu$ sector which arise from the $Y_\ell^{\mu\tau}$ and $Y_\ell^{\tau\mu}$ couplings. For simplicity, we will take $Y_\ell^{\mu\tau}=Y_\ell^{\tau\mu}$ to be real. If other components of $Y_\ell$, such as $Y_\ell^{\tau\tau}$, are non-vanishing, they will dilute the $\tau$-$\mu$ flavor violating branching ratios of the heavy scalars. The branching ratios of the heavy scalars are shown in Fig.~\ref{fig:br}.

%
\subsection{LHC Searches for Heavy Scalars}\label{subsec:LHCcharged}
%
The heavy scalars have been extensively searched for at the LHC. In particular CMS has been searching for the s-channel production of the heavy Higgs decaying into a tau-pair~\cite{Sirunyan:2018zut} and $W$-pair~\cite{CMS:2019kjn}. The former analysis covers the mass region $m_H>90$ GeV while the latter covers $m_H>200$ GeV. Moreover, CMS has performed a search for a pseudoscalar decaying into a pair of $hZ$~\cite{CMS:1900sig}. All three analyses are based on 35.9 fb$^{-1}$ of the LHC Run II data.

The charged Higgs can be searched for in the decay channel $H^+\to t\bar b$ in the case of a heavy charged Higgs, $m_{H^+}>m_t+m_b$, and $H^+\to \tau\nu$ in the case of a light charged Higgs. The former search has been performed by ATLAS~\cite{Aaboud:2018cwk} using 36.1 fb$^{-1}$ of the 13 TeV LHC data while the latter search has been analyzed by CMS~\cite{Sirunyan:2019hkq} using 35.9 fb$^{-1}$ of data. 

We will collectively refer to the constraints placed by the above searches on the parameter space of the Type-III 2HDM as the non-LFV bounds. In the next section, we will only show the strongest constraint among all the analyses above.

\section{LHC Constraints on $\tau$-$\mu$ LFV} \label{sec:param}
\begin{figure}[t!]
\centering \includegraphics[width=0.45\textwidth]{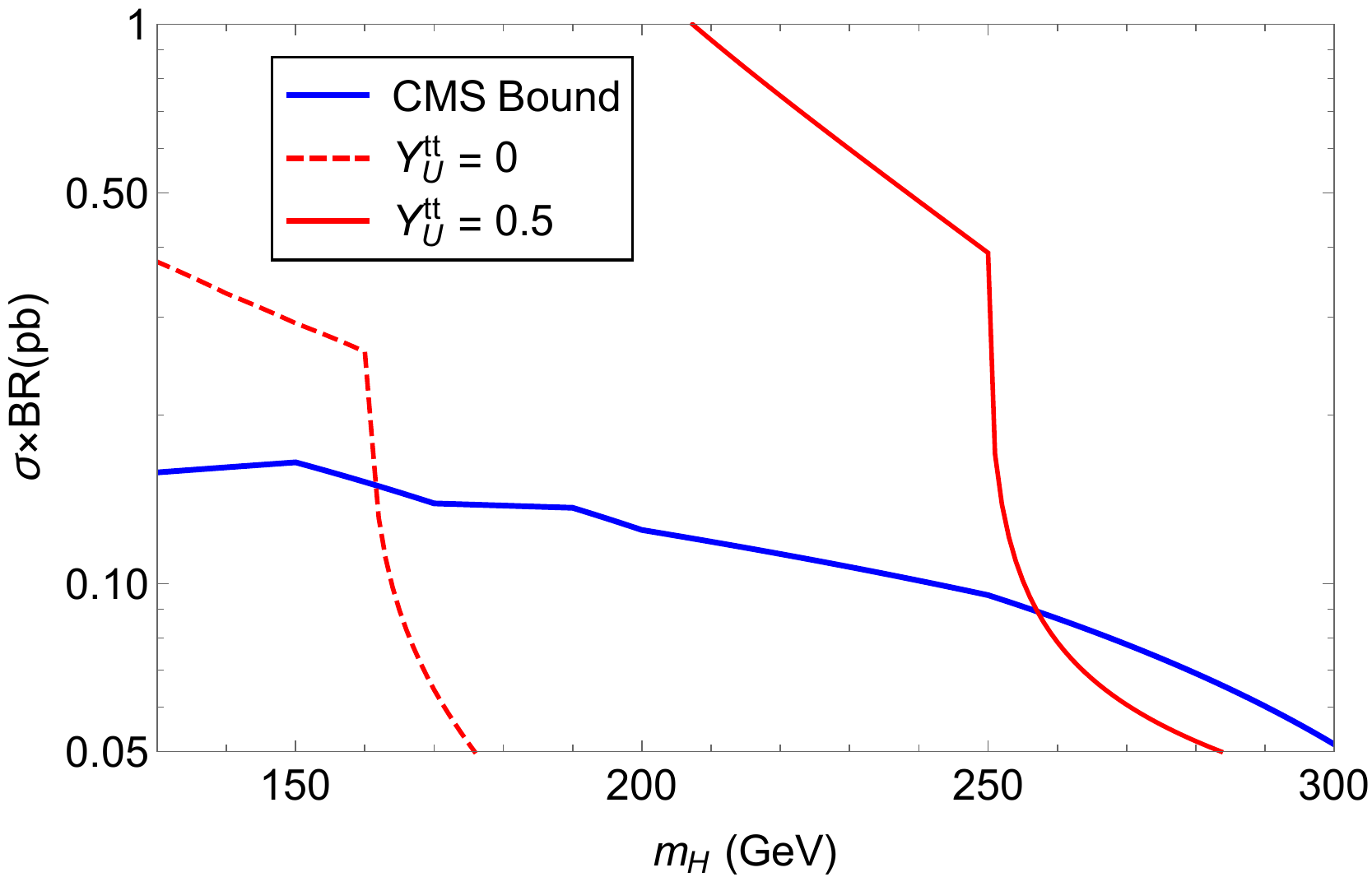}
\caption{The $\sigma\times$BR($\tau\mu$) of the heavy Higgs in the case $\sin \alpha = 0.1$ and $\lambda_{Hhh} = 0.5$. The plot shows two cases: the small production case ($Y_U^{tt} = 0$, red dashed line) and the large production case ($Y_U^{tt} = 0.5$, red solid line). For comparison, the recast CMS bound discussed in Sec.~\ref{sec:LHC} is shown in blue.}
\label{fig:boundscompare}
\end{figure}

%
In this section, we will present the viable parameter space of the Type-III 2HDM with LFV in the $\tau$-$\mu$ sector with respect to collider searches.
We start by computing the $\sigma\times$BR($\tau\mu$) of the heavy Higgs as a function of its mass and comparing it against the recast CMS analysis.
In Fig.~\ref{fig:boundscompare}, the $\sigma\times$BR($\tau\mu$) is shown for both the small production case ($Y_U^{tt} = 0$) and the large production case ($Y_U^{tt} = 0.5$). In making these plots we take $\sin \alpha = 0.1$ and $\lambda_{Hhh} = 0.5$. 
From the plot, we see that in the small production case, the CMS bounds become irrelevant once the phase space for $H\to W^+W^-$ opens, i.e. $m_H \gtrsim 160$ GeV. In the strong production case, the bounds can be extended beyond the $W^+W^-$ threshold. In our particular benchmark case, the bound extends to around $m_H \sim 2 m_h$, where the decay $H\to hh$ opens. However the bound could be extended further if the coupling $\lambda_{Hhh}$ is smaller, which in turn enhances the $\tau$-$\mu$ branching ratio of the heavy Higgs.

In the rest of this section, we will further analyze the implication of the CMS heavy Higgs $\tau$-$\mu$ LFV search on the Type-III 2HDM. We will also study the interplay between the light Higgs LFV search, the heavy Higgs LFV search and the heavy scalar searches in constraining the model parameter space.

\subsection{Large Production Case}\label{subsec:large}
%
In this subsection, we analyze the most optimistic scenario in which the heavy scalars production is enhanced, while their branching fractions to $\tau$-$\mu$ are not diluted by other decay channels. This scenario corresponds to keeping only $Y_U^{tt}$ and $Y_\ell^{\tau\mu}=Y_\ell^{\mu\tau}$.

The viable parameter space of the model depends on the scalar mass spectrum. To simplify the discussion, we consider three representative cases: 1) the $H$ is the lightest among the heavy scalars, 2) the $A$ is the lightest amongst the heavy scalars, and 3) all heavy scalars are degenerated. For the first two cases, we will take the remaining two heavy scalars to be degenerated in mass and are 100 GeV heavier than the lightest one. In all the above representative cases, we have verified that the scalar spectrum is consistent with the electroweak $S$ and $T$ parameters over the mass range considered in our analyses.

%
\subsubsection{Lightest $H$}\label{subsec:lightestH}
%

\begin{figure*}[thbp]
     \begin{center}
         \includegraphics[width=0.4\textwidth]{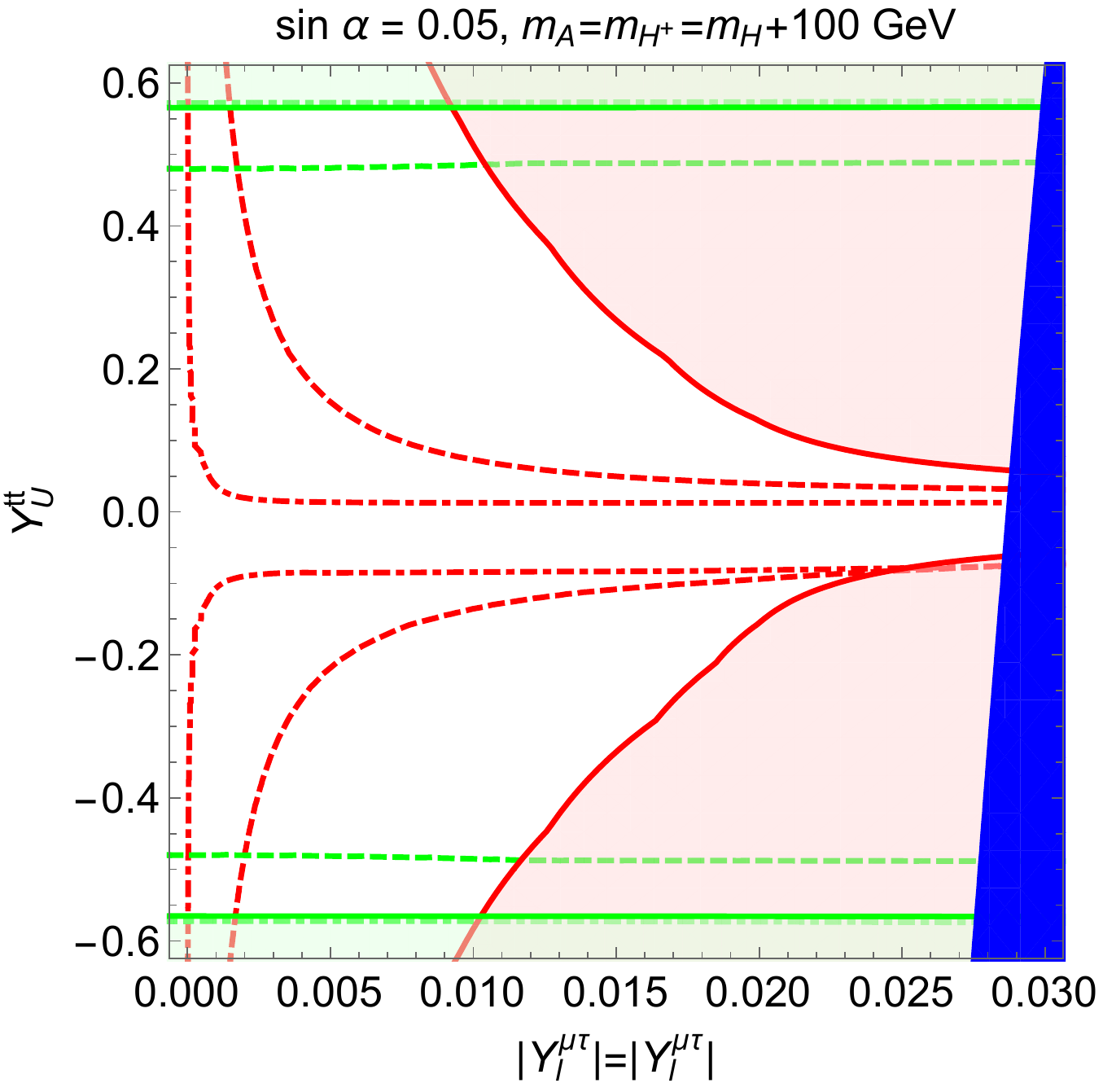}\qquad
         \includegraphics[width=0.4\textwidth]{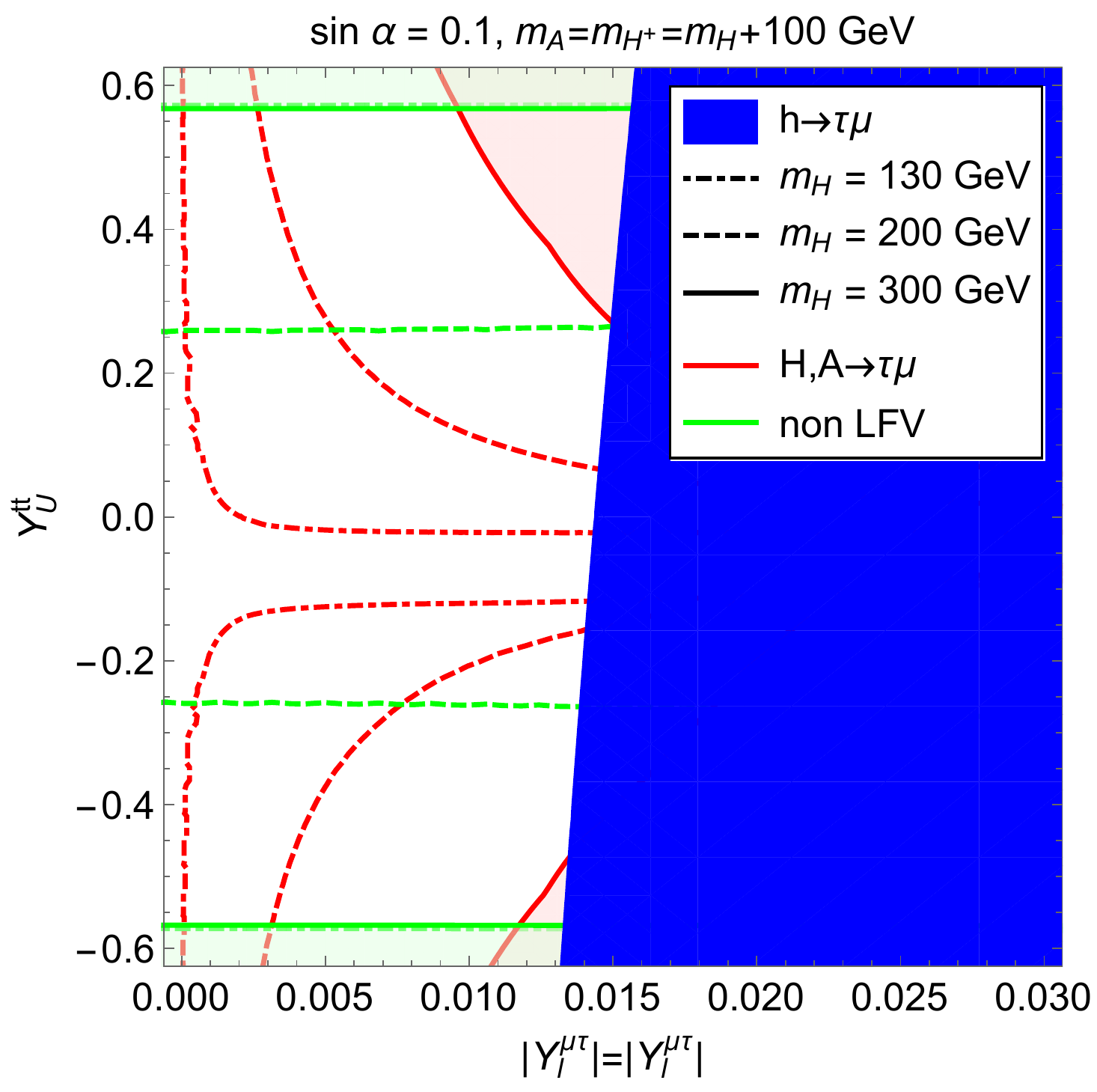}\qquad
     \end{center}
        \caption{The excluded parameter space in the case of $\sin\alpha = 0.05, 0.1$, $m_A = m_{H^{+}}= m_H + 100$ GeV, $\lambda_{Hhh} = 0.5$ and  $m_H=130$, 200 and 300 GeV. The blue region is excluded by the light Higgs LFV search~\cite{Sirunyan:2017xzt}. 
        The  allowed regions from the CMS heavy Higgs LFV search~\cite{Sirunyan:2019shc} are denoted by the area between the two red lines for each corresponding $m_H$. 
        The allowed regions from the non LFV heavy scalar searches are denoted by the area between the green lines.
        The dash-dotted, dashed and solid lines represent the boundaries of the excluded regions for $m_H$ = 130, 200 and 300 GeV respectively.
        }
        \label{fig:lightH}
\end{figure*}

We will first discuss the case where the scalar $H$ is the lightest with $m_{H^{+}} = m_A = m_H + 100$ GeV. The exclusion regions in the $Y_\ell^{\mu\tau}$-$Y_U^{tt}$ plane for $\sin\alpha=0.05$ and 0.1, together with $m_H=130$, 200 and 300 GeV, are shown in Fig.~\ref{fig:lightH}. Comparing the left and the right plots, we see that the $h$ LFV bounds become less relevant as $\sin\alpha$ gets smaller.

We also note that, for a fixed $\sin\alpha$, the bounds from the heavy Higgs LFV search get stronger as $m_H$ becomes lighter. In the case of $m_H = 130$ GeV, the bounds from the heavy Higgs LFV search exclude most of the parameter space, except when either $Y_\ell^{\mu\tau}$ becomes very small or the combination of $\sin\alpha$ and $Y_U^{tt}$ conspires to have the $H$ production cross-section vanished. Only in the case of a very heavy $H$, the constraints from the LFV decay of the light Higgs become relevant in most of the parameter space. 

The non-LFV constraints depends heavily on the mass. For $m_H = 130$ GeV ($m_{H^+}=m_A$ = 230 GeV), the most relevant constraint is the charged Higgs search in the $H^+\to t \bar b$  channel. For $m_H = 200$ GeV ($m_{H^+}=m_A = 300$ GeV), the search for pseudoscalar decaying into $hZ$ provides a more stringent constraint on the parameter space compared to the $H^+\to t \bar b$ search. Finally, for $m_H = 300$ GeV ($m_{H^+}=m_A = 400$ GeV), the pseudoscalar decays mostly into $t \bar t$, diluting the $A\to hZ$ signal. As a result, the $H^+ \to t \bar b$ search is again becoming the most constraining search.

%
\subsubsection{Lightest $A$}\label{subsec:lightestA}
\begin{figure*}[tbp]
     \begin{center}
         \includegraphics[width=0.4\textwidth]{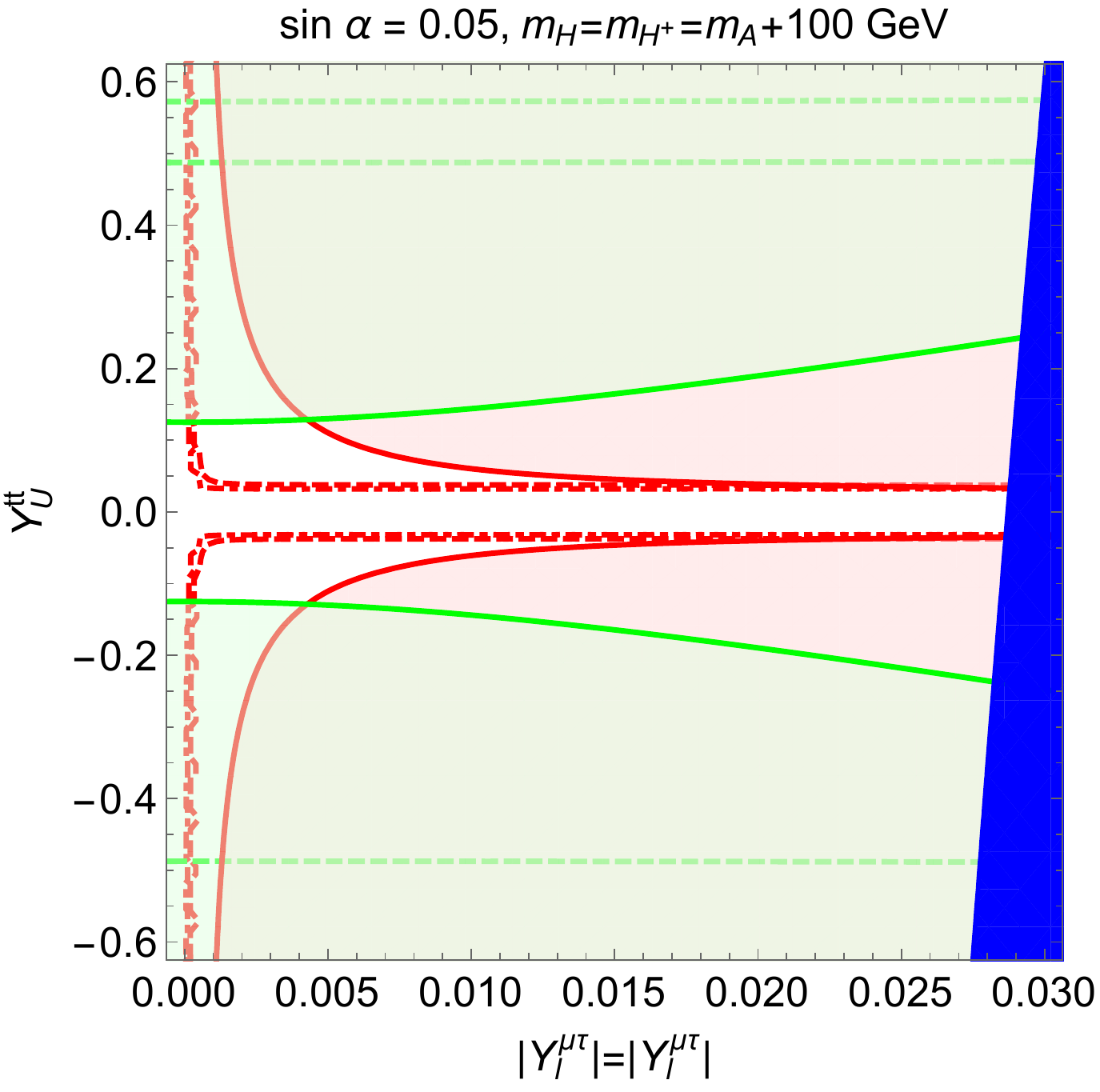}\qquad
         \includegraphics[width=0.4\textwidth]{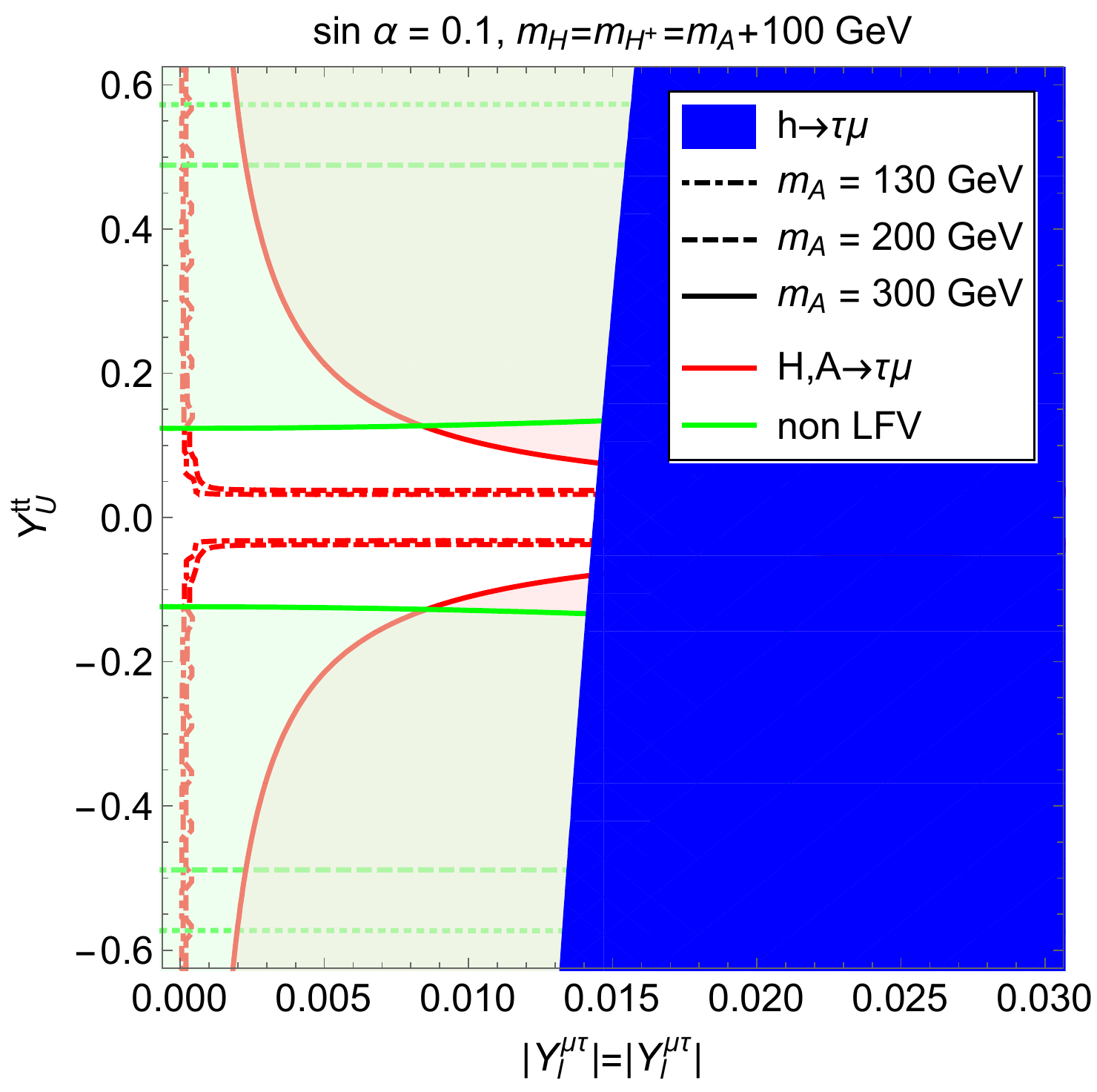}\qquad
     \end{center}
        \caption{The excluded parameter space in the case of $\sin\alpha = 0.05, 0.1$, $m_H = m_{H^{+}}= m_A + 100 $ GeV, $\lambda_{Hhh} = 0.5$ and $m_A=130$, 200 and 300 GeV. The blue region is excluded by the light Higgs LFV search~\cite{Sirunyan:2017xzt}. 
        The  allowed regions from the CMS heavy Higgs LFV search~\cite{Sirunyan:2019shc} are denoted by the area between the two red lines for each corresponding $m_H$. 
        The allowed regions from the non LFV heavy scalar searches are denoted by the area between the green lines.
        The dash-dotted, dashed and solid lines represent the boundaries of the excluded regions for $m_H$ = 130, 200 and 300 GeV respectively.}
        \label{fig:lightA}
\end{figure*}

%
We turn next to the case in which the pseudoscalar $A$ is the lightest with $m_{H^{+}} = m_H = m_A + 100$ GeV. 
The exclusion plots for this case are shown in Fig.~\ref{fig:lightA}. We see that the pseudoscalar LFV searches generically provide better constraints compared to the $h$ LFV searches.
Several comments on the bounds are in order. 
For the pseudoscalar mass below the $hZ$ threshold, the $A$ decays mostly into $\tau\mu$. Thus the pseudoscalar LFV bounds for $m_A=130$ and 200 GeV are stronger than the corresponding $H$ LFV bound with the same mass. For $m_A=300$ GeV, the $A\to hZ$ is open, diluting the $A\to\tau\mu$ signal. However, for our scenario, the branching ratio $A\to\tau\mu$ with $m_A=300$ GeV is still larger than the corresponding branching ratio for the $H$ with the same mass. As a result, the pseudoscalar LFV search provides a better constraint in this case as well. Lastly, we note that the production cross-section of the the pseudoscalar does not depend on the mixing angle $\sin\alpha$. Therefore the heavy Higgs LFV search bounds, shown in the red lines on the left and the right plots of Fig.~\ref{fig:lightA}, do not change significantly as we vary the value of the mixing angle.

We conclude this subsection by discussing the non-LFV bounds. For $m_A\leq 200$ GeV, the most relevant constraint comes from the charged Higgs search in the $t\bar b$ channel.
For $m_A = 300$ GeV, the decay channel $A\to hZ$ opens. Hence the search for $A\to hZ$ provides the most stringent non-LFV constraint in this case.

%
\subsubsection{Degenerate Spectrum}\label{subsec:degenerate}
\begin{figure*}[tbp]
     \begin{center}
         \includegraphics[width=0.4\textwidth]{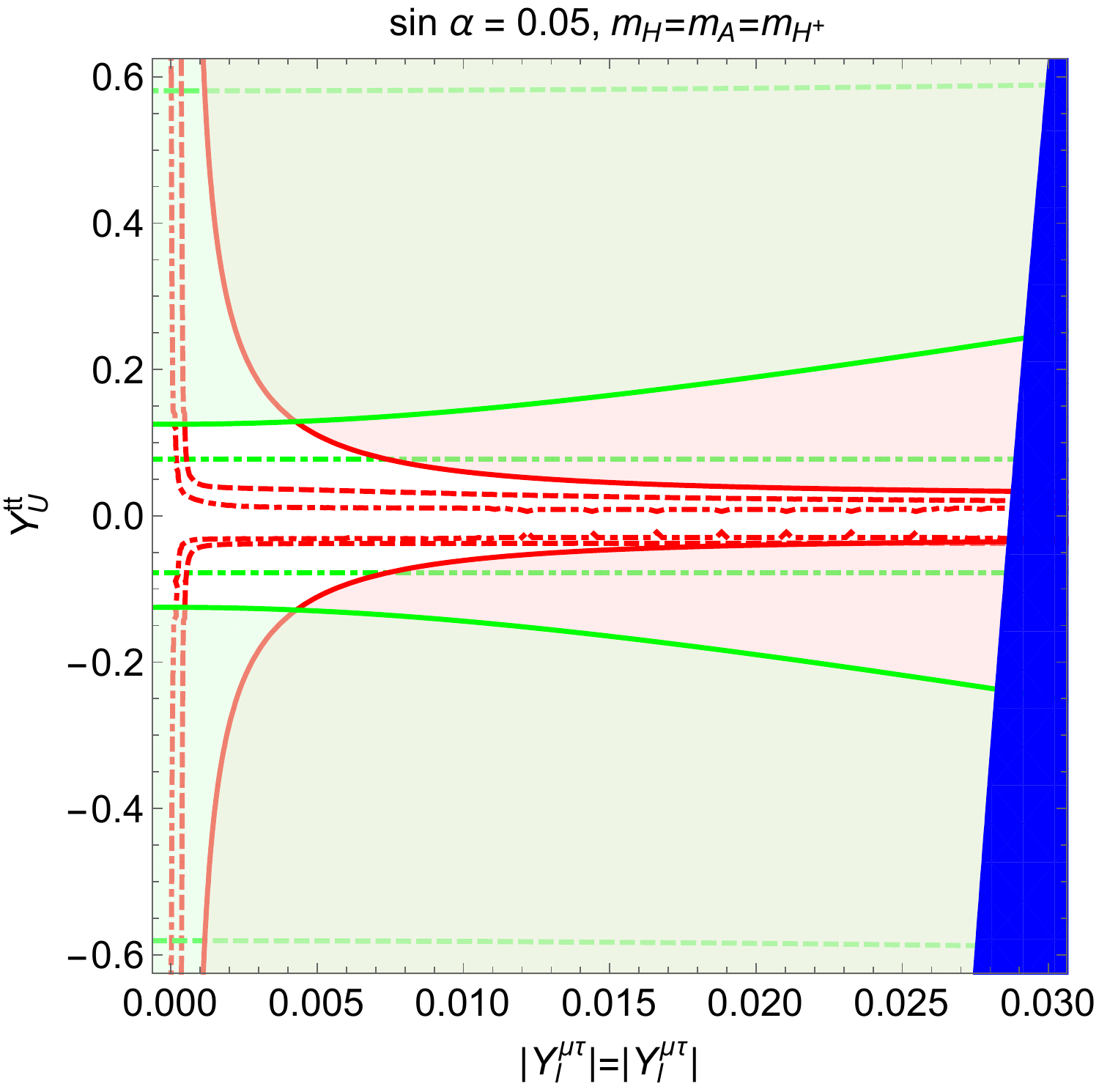}\qquad
         \includegraphics[width=0.4\textwidth]{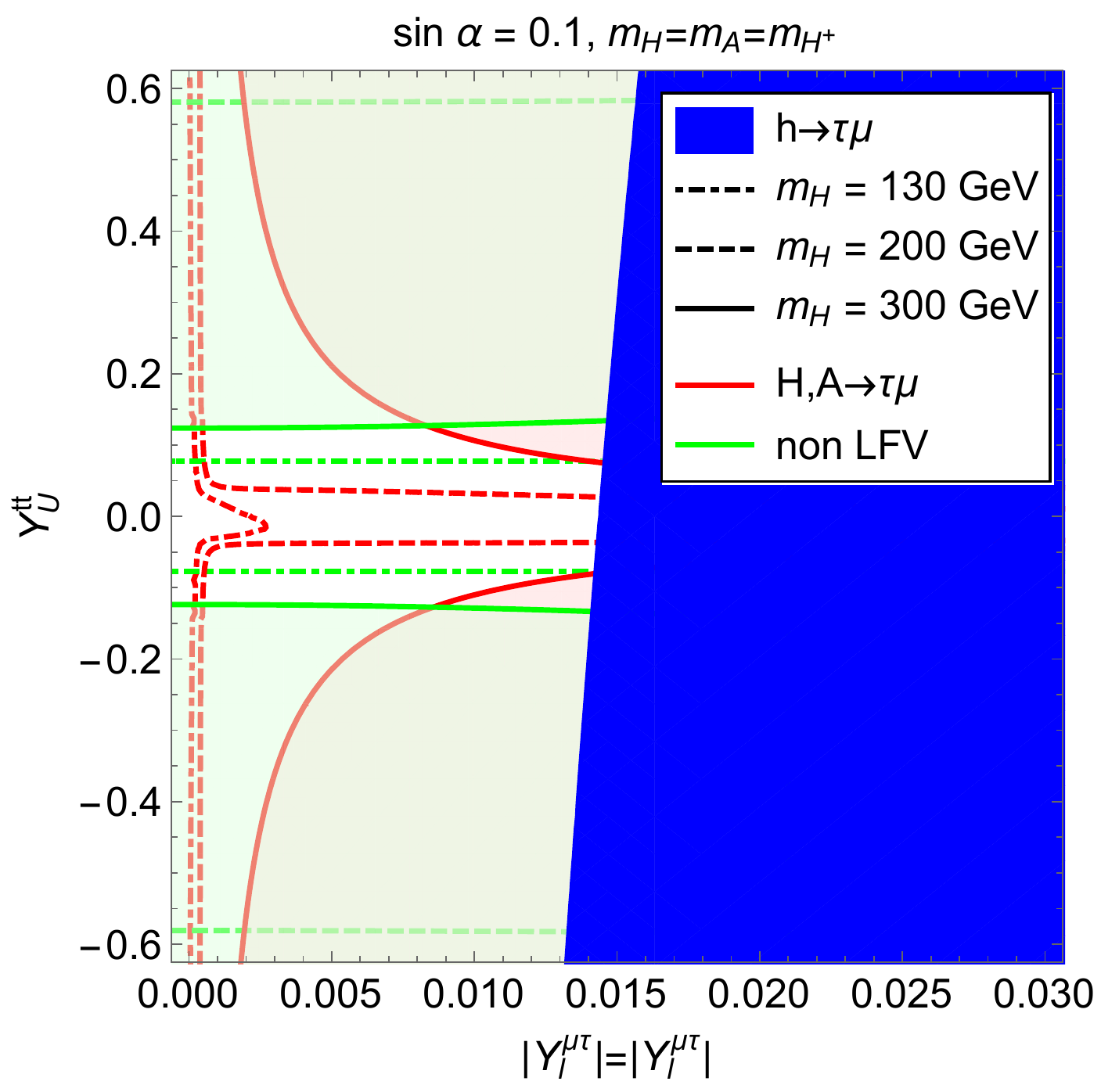}
     \end{center}
        \caption{The excluded parameter space in the case of $\sin\alpha = 0.05, 0.1$, $m_H = m_A = m_{H^{+}}$, $\lambda_{Hhh} = 0.5$ and $m_H=130$, 200 and 300 GeV. The blue region is excluded by the light Higgs LFV search~\cite{Sirunyan:2017xzt}. 
        The  allowed regions from the CMS heavy Higgs LFV search~\cite{Sirunyan:2019shc} are denoted by the area between the two red lines for each corresponding $m_H$. 
        The allowed regions from the non LFV heavy scalar searches are denoted by the area between the green lines.
        The dash-dotted, dashed and solid lines represent the boundaries of the excluded regions for $m_H$ = 130, 200 and 300 GeV respectively.}
        \label{fig:degenerate}
\end{figure*}

%
In this subsection we consider the case where all the heavy scalars are degenerated. 
The viable parameter space in this case is shown in Fig.~\ref{fig:degenerate}. 
We see that the bounds from the heavy Higgs LFV search become stronger compared to the two previous benchmarks because now the two heavy neutral scalars are equally light and can be produced with larger cross-section. Similarly the bounds from the non-LFV heavy scalar searches also become stronger. On the other hand, the bounds from the light Higgs LFV search are independent of the heavy scalar spectrum, hence they do not change from the previous benchmarks. 
We note that in the degenerate case, the heavy scalar LFV searches generically provide more significant constraints compared to the light Higgs LFV search. For our particular benchmark, the bounds are mostly driven by the pseudoscalar searches since its LFV branching ratio is bigger than that of the heavy Higgs. 
It is worth noting that for sufficiently light heavy scalars, such as $m_H=m_A=130$ GeV, with a large enough mixing, the heavy scalars LFV search places a very strong bound on the $Y_\ell^{\tau\mu}$ and $Y_\ell^{\mu\tau}$. For $m_H=m_A=130$ GeV with $\sin\alpha=0.1$, $Y_\ell^{\tau\mu}$ and $Y_\ell^{\mu\tau}$ are constrained to be less than 0.003, see the right plot in Fig.~\ref{fig:degenerate}.

Let us end this subsection with a few remarks regarding the non-LFV searches. The most relevant constraint in this case depends strongly the heavy scalar mass. For the case $m_H = 130$ GeV, the most constraining bound comes from the charged Higgs search in the $\tau \nu $ channel. 
For $m_H = 200$ GeV, the charged Higgs search in the $t \bar b$ channel provides the strongest constraint. Finally, in the case $m_H = 300$ GeV, the phase space for $A \rightarrow h Z$ opens. Hence the $A\to hZ$ search provides the strongest bounds.

%
\subsection{Small Production Case}\label{subsec:small}
%
In the previous section, we have discussed the case where the production of the heavy scalar is enhanced by the presence of $Y_U^{tt}$. In that optimistic scenario, we see that the CMS heavy Higgs LFV search can put bounds on a large section of the parameter space unconstrained by the light Higgs LFV search. In this section, we discuss the relevance of the heavy Higgs LFV search in the pessimistic scenario where $Y_U^{tt}$ is absent. In this scenario, only the heavy Higgs can be produced via the neutral CP-even scalars mixing.  
We introduce two benchmarks, the first is the less pessimistic case where the $H\to\tau\mu$ branching fraction does not get diluted. We will refer to this benchmark as the mixing production case. The second benchmark represents the most pessimistic case where the $H\to\tau\mu$ branching fraction is diluted by the presence of $Y_{\ell}^{\tau\tau}$. We will refer to this benchmark as the mixing production with $Y_\ell^{\tau\tau}$ case.
%
\subsubsection{Mixing Production}\label{subsec:noastrongprod}
%
In this subsection, we consider the case in which all $Y$'s except $Y_\ell^{\tau\mu}$ and $Y_\ell^{\mu\tau}$ are vanishing. The scalar $H$ can only be produced via the mixing in the neutral component of doublet $\Phi_1$. Its production cross-section is controlled only by the mixing angle $\sin \alpha$ and the mass $m_H$. 
As for the pseudoscalar, since $Y_U^{tt}$ is zero, it cannot be produced via gluon fusion and hence does not contribute to the constraints on the parameter space. 
Moreover, non-LFV heavy Higgs searches discussed in Sec.~\ref{subsec:LHCcharged} require that $Y_U^{tt} \neq 0$. Therefore, they also do not provide any constraints in this scenario. 

\begin{figure}[ht!]
\centering
         \includegraphics[width=0.4\textwidth]{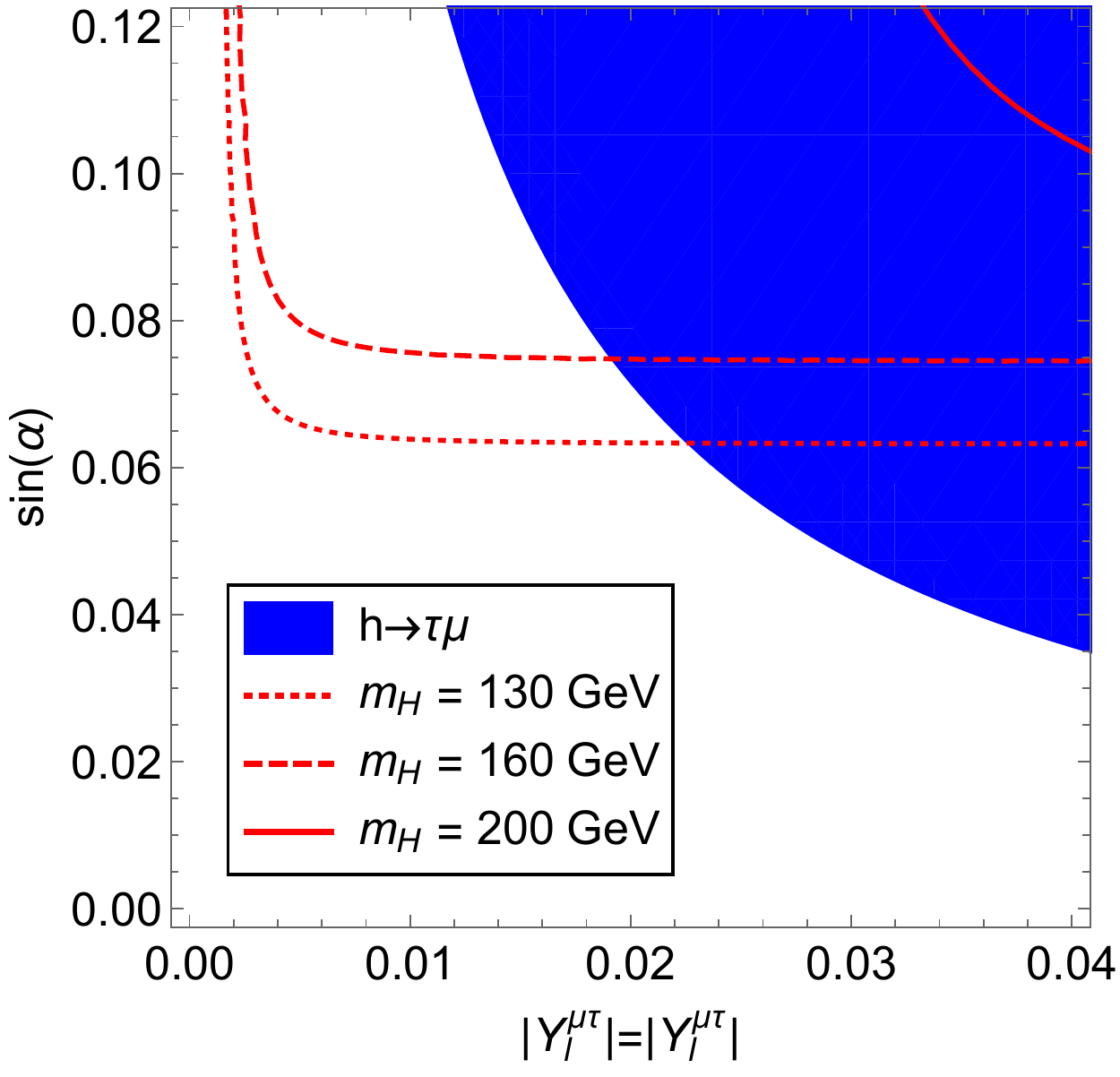}
           \caption{The excluded parameter space for $m_H=130$, 160 and 200 GeV with all $Y$'s taken to be zero expect $Y_\ell^{\mu\tau}$. The blue region is the excluded by the light Higgs LFV search~\cite{Sirunyan:2017xzt}. 
        The areas above the red lines are excluded by the CMS heavy Higgs LFV search~\cite{Sirunyan:2019shc}. 
        The dash-dotted, dashed and solid lines represent the boundaries of the excluded regions for $m_H$ = 130, 160 and 200 GeV respectively.}
        \label{fig:mixingonly}
\end{figure}

The excluded parameter space is shown in Fig.~\ref{fig:mixingonly}. The bounds are mostly insensitive to the heavy scalar mass spectrum, as long as the $H$ does not decay into other scalars. From the plot we can see that only in the case of $m_H \lesssim 2m_W$, the heavy Higgs LFV search becomes relevant. For $m_H \gtrsim 2 m_W$, the heavy Higgs decays dominantly into $WW$, diluting the LFV signal and weakening its bound. Hence the heavy Higgs LFV search is only relevant for the case of $m_H \lesssim 2 m_W$, which was not considered in the official CMS analysis~\cite{Sirunyan:2019shc}. 

%
%
\subsubsection{Mixing Production With $Y_\ell^{\tau\tau}$}\label{subsec:ytautau}
%

In this subsection, we suppress the $H\to\tau\mu$ branching fraction by introducing an extra lepton Yukawa $Y_\ell^{\tau\tau}$. 
There are several constraints when the $Y_\ell^{\tau\tau}$ is present. The $Y_\ell^{\tau\tau}$ is constrained by the signal strength measurement of $h \rightarrow \tau^+ \tau^-$. The latest measurements by both ATLAS and CMS give $\mu = 0.82 \pm 0.16$~\cite{Sirunyan:2018koj,*Aad:2019mbh}. 
Moreover, the presence of $Y_\ell^{\tau\tau}$ will induce a $\tau \rightarrow \mu \gamma$ decay. However, we find that the constraint from $\tau\to\mu\gamma$ decay is negligible compared to the LFV collider bounds for our benchmark scenarios.

\begin{figure}[t]
\centering
         \includegraphics[width=0.4\textwidth]{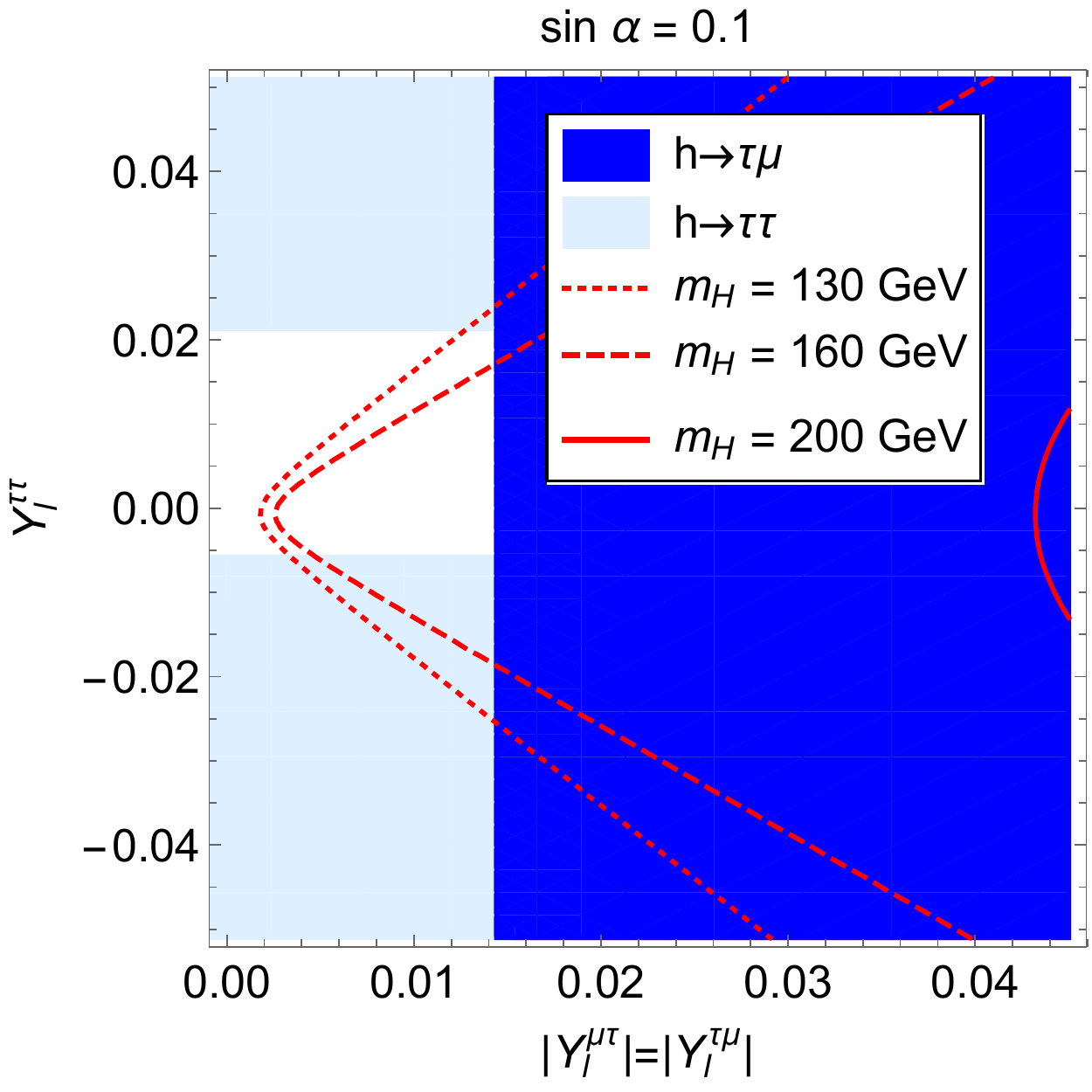}
        \caption{The excluded parameter space for $m_H=130$, 160 and 200 GeV with nonzero $Y_\ell^{\tau\tau}$ and $Y_\ell^{\tau\mu}=Y_\ell^{\mu\tau}$ and $\sin\alpha = 0.1$. The blue region is the excluded by the light Higgs LFVsearch~\cite{Sirunyan:2017xzt}. The red region represents the boundary of excluded regions from the CMS heavy Higgs LFV search~\cite{Sirunyan:2019shc}. All the area on the right of the respective red lines are excluded. The light blue region is excluded by the signal strength measurement of light Higgs decay to $\tau^+\tau^-$.}
        \label{fig:yttsa01}
\end{figure}

Fig.~\ref{fig:yttsa01} shows the bounds on the model parameter space in the $Y_\ell^{\mu\tau}$ and $Y_\ell^{\tau\tau}$ plane. The plot shows that even in this case, the heavy Higgs LFV search can still provide substantial bounds, provided that the value of $\sin\alpha$ is large enough and the heavy Higgs mass is below the $WW$ threshold. In this case, the $H\to\tau\mu$ branching fraction is only suppressed by the decay into $\tau^+\tau^-$ which, while significant, is still on the same order as the $H\to\tau\mu$ decay.

\section{Conclusions} \label{sec:conclusions}

In this work we have studied the significance of the recent CMS heavy Higgs LFV search~\cite{Sirunyan:2019shc}. First, we recast the CMS analysis to obtain the $\sigma\times$BR($\tau\mu$) bounds for the heavy Higgs mass between 130 GeV and 450 GeV. Then we study the implication of the bounds in the context of the Type-III 2HDM. We consider the most optimistic and pessimistic scenarios for probing the $\tau$-$\mu$ LFV. In the optimistic scenarios, the  production cross-sections of the heavy scalars are large because of the presence of $Y_U^{tt}$. 
In this case, we have shown that the official CMS heavy Higgs LFV search generically extends the exclusion regions of the model parameter space of the light Higgs LFV search. Moreover, the bounds from the heavy Higgs LFV search are especially strong when the heavy scalars are light. 
In the pessimistic scenarios, in which the heavy Higgs is produced by the mixing with the light Higgs, we found that the heavy Higgs LFV search still plays an important role in probing the parameter space of the model. However, in this case the heavy Higgs LFV search is only relevant for $m_H \lesssim 2m_W$, i.e., below the $W^+W^-$ decay threshold, see Fig.~\ref{fig:mixingonly}. Unfortunately, this low mass region does not get covered by the official CMS search which only considers $m_H > 200$ GeV. Thus, we suggest CMS to consider extending their heavy Higgs mass region to below the $W^+W^-$ threshold in their future analysis.


\bigskip
\paragraph*{Acknowledgments:}
%
The work of RP is supported by the Parahyangan Catholic University under grant no.
III/LPPM/2019-01/42-8. The work of JJ is supported by Indonesian Institute of Sciences with grant no. B-288/IPT/HK.02/II/2019. The work of PU has been supported in part by the Thailand Research Fund under contract no.~MRG6280186, and the Faculty of Science, Srinakharinwirot University under grant no.~222/2562 and 498/2562.

\appendix

\bibliography{lit1} \bibliographystyle{apsrev4-1}

\end{document}